\documentclass[aps,prd,twocolumn,superscriptaddress,preprintnumbers,floatfix,nofootinbib,notitlepage,eqsecnum,showpacs]{revtex4}

\usepackage{graphicx}
\usepackage{latexsym}
\usepackage{amsmath,amssymb}
\usepackage{amsfonts}
\usepackage{longtable,booktabs}

\newcommand{\ev}[1]{\left \langle #1 \right  \rangle}
\newcommand{\MB}[2]{ I_{#1} \left ( #2 \right  )}
\usepackage{placeins}
\usepackage{hyperref}
\hypersetup{
    colorlinks=true,     
    linkcolor=blue,      
    citecolor=blue,      
    filecolor=blue,      
    urlcolor=blue        
}

\begin{document}
\title{Benchmark results in the 2D lattice Thirring model with a chemical potential}
\author{Venkitesh Ayyar}
\affiliation{Department of Physics, University of Colorado, Boulder, CO 80309, USA}
\author{Shailesh Chandrasekharan}
\affiliation{Department of Physics, Box 90305, Duke University, Durham, NC 27708, USA}
\author{Jarno Rantaharju}
\affiliation{Department of Physics, Box 90305, Duke University, Durham, NC 27708, USA}

\begin{abstract}
We study the two dimensional lattice Thirring model in the presence of a fermion chemical potential. Our model is asymptotically free, contains massive fermions that mimic a baryon and light bosons that mimic pions. Hence it is a useful toy model for QCD, especially since it too suffers from a sign problem in the auxiliary field formulation in the presence of a fermion chemical potential. In this work we formulate the model in both the world line and fermion-bag representations and show that the sign problem can be completely eliminated with open boundary conditions when the fermions are massless. Hence we are able accurately compute a variety of interesting quantities in the model, and these results could provide benchmarks for other methods that are being developed to solve the sign problem in QCD.
\end{abstract}


\maketitle
\section{Introduction}

Traditional lattice calculations of quantum field theories often encounter sign problems in the presence of a chemical potential. An excellent example is Quantum Chromodynamics (QCD), where it is impossible to accurately compute quantities at a non-zero baryon density especially at low temperatures \cite{deForcrand:2010ys}. Over the past decade, ideas like the complex Langevin approach \cite{Seiler:2017wvd} and the Leftchetz thimble approach \cite{Cristoforetti:2012su} have been proposed as potential solutions to sign problems including QCD. When these methods are tested on simple models where exact results are available \cite{Mukherjee:2013aga,Tanizaki:2015rda,Fujii:2015vha}, we not only find potential pitfalls of the methods but also learn new directions to avoid them \cite{Nishimura:2015pba, Bloch:2016jwt,Aarts:2017vrv}. While these ideas have also been able to capture some of the qualitative features of more complex field theories \cite{Aarts:2016qrv,Mukherjee:2014hsa}, in these cases the numerical results are not always compared with benchmark calculations obtained with other methods where the errors can be controlled. An exception to this has been studies of bosonic field theories at finite densities where a controlled Monte Carlo algorithm in the world line representation free of sign problems is available \cite{Cristoforetti:2013wha,Aarts:2010aq}. Producing such benchmark calculations that truly test the method, especially in fermionic quantum field theories with a sign problem and similar to QCD in other aspects, would be helpful and is the main motivation behind our work.

\begin{figure}[b]
\includegraphics[width=0.8\linewidth]{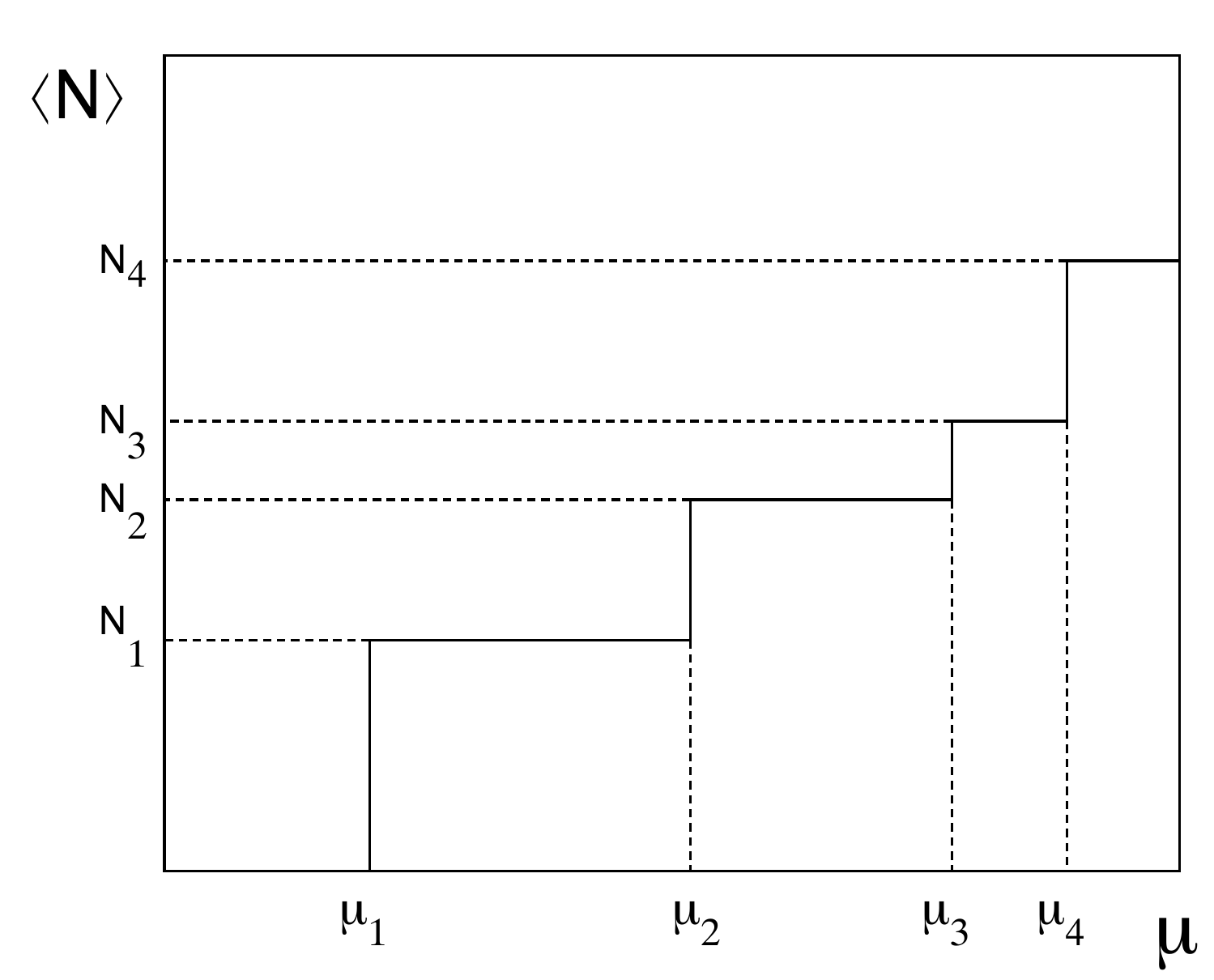}
\caption{A schematic plot of the particle number as a function of chemical potential for a fixed spatial size. We propose that the values of $\mu_i$ and the corresponding $N_i$'s are easily calculable and can be used as benchmark quantities to validate a method that claims to solve a sign problem.}
\label{fig1}
\end{figure}

Recently the Lefshetz thimble program got a boost when it was shown that it may be possible to use holomorphic flow in complex field space to sample multiple thimbles rather than perform calculations on a single thimble as was done in the past \cite{Alexandru:2015sua}. The focus has also turned to lattice Thirring models as a prototype example of the physics of QCD \cite{Alexandru:2015sua,Alexandru:2016ejd}. This model has also been studied earlier in higher dimensions using stochastic quantization \cite{Pawlowski:2013gag}. Also, the recent work has computed the average fermion number $\langle N\rangle$ on small but fixed spatial size $L_X$ as a function of the chemical potential, which is much more sensitive to the  important physical scales in the problem, as compared to local densities on large space-time lattices. As shown schematically in Fig.~\ref{fig1}, at low temperatures (or large $L_t$) the plot of $\langle N\rangle$ as a function of the chemical potential $\mu$ is expected to show a series of jumps at critical values of the chemical potential, say $\mu_1,\mu_2,...$ where the average particle number jumps to $N_1,N_2,...$. The values of $\mu_i$'s and $N_i$'s are related to the physical scales of the problem like binding energies and scattering lengths and should become harder to compute due to sign problems, especially when $\mu_i$ and $N_i$ become large. Encouraged by the fact that some of these quantitative features may be within reach, recently the efforts have turned towards speeding up the calculations on larger lattices using machine learning algorithms \cite{Alexandru:2017czx}. It would indeed be exciting if this program is successful. 

The motivation for our work is to help this program by accurately computing the $\mu_i$'s and $N_i$'s for a specific two dimensional lattice Thirring model constructed with staggered fermions. Our model is asymptotically free and a continuum limit can be defined at zero coupling. At nonzero couplings (finite lattice spacing) the fermion in the theory is massive and mimics a baryon, while bosonic excitations made with fermion-antifermion pairs are massless and mimic pions. Thus, the similarities of our model with QCD is striking. Of course the ground state does not break any symmetries and the pions are not really Goldstone bosons as was explained by Witten long ago \cite{Witten:1978qu}, but the fermion mass generation is dynamical like in QCD and from the point of view of sign problems the bosons being lighter than the fermion is also similar to QCD. Interestingly, we can solve the model both in the fermion world line method and the fermion bag approach. In the world line approach we argue in this work that the sign problem is absent with open boundary conditions and zero fermion mass. Thus, in this limit we are able to study large lattices and can accurately compute the the critical $\mu_i$'s and $N_i$'s. These could provide helpful benchmark to test new ideas that claim to solve sign problems in problems similar to QCD.

Our paper is organized as follows. In section \ref{model} we discuss the model we study and the various types of representations that can be used to solve it. In particular we show why the model in the massless limit with open boundary conditions has no sign problem in the worldline formulation.  In section \ref{mcmethods} we discuss our Monte Carlo methods, especially the worm algorithm to update the worldline representation and the fermion bag algorithm. In section \ref{results} we discuss the results we have obtained. In particular we define the observables we measure and discuss our results in a variety of parameter ranges. We present our conclusions in section \ref{conclusions}.

\section{The Model} \label{model}

The lattice action of the model we study is given by
\begin{equation}
S = \sum_{x,y} \overline{\chi}_x (M_{x,y} + m\delta_{x,y}) \chi_y  +  U \sum_{x,\nu} \overline{\chi}_x \chi_{x+\nu} \overline{\chi}_{x+\nu} \chi_x.
\end{equation}
where the the matrix $M$ is the massless staggered fermion matrix defined as 
\begin{equation}
M_{x,y} = \sum_{\nu} \frac {\eta_{x,\nu}}{2} \left (  e^{\mu \delta_{\nu,0} }  \delta_{x+\nu,y} - e^{ - \mu \delta_{\nu,0 } } \delta_{x,y+\nu} \right ),
\end{equation}
where $\mu$ is the chemical potential, $m$ is the fermion mass and $\eta_{x,\mu}$ are the usual staggered phase factors ($\eta_{x,0} = 1$ and $\eta_{x,1} = (-1)^{x_1}$). The four-fermion coupling $U$ can be interpreted as a current-current interaction on neighboring sites and hence the name ``lattice Thirring model''. When $m=0$ the model contains the well known $U(1)$ chiral symmetry of staggered fermions. In the discussion below $L_X$ denotes the number of spatial sites and $L_T$ denotes the number of temporal sites in our two dimensional square lattice. Further, we always use anti-periodic boundary conditions in time, but study the effects of periodic, anti-periodic and open boundary conditions in space.

This model has a long history and has been studied extensively in three space-time dimensions in the auxiliary field formulation \cite{DelDebbio:1995zc,DelDebbio:1997dv} and the fermion bag approach \cite{Chandrasekharan:2009wc,Chandrasekharan:2011mn}. In three dimensions the model with $m=0$ has two phases: a weak coupling phase with massless fermions and a strong coupling phase with spontaneously broken $U(1)$ chiral symmetry, massive fermions and light pions. These phases are separated by a second order critical point, whose properties were studied in the earlier work. In two dimensions this critical point moves to the origin and the massless weak coupling phase disappears. Further, since a continuous chiral symmetry cannot break in two dimensions, the massive fermion phase becomes critical. Thus, the two dimensional model contains massive fermions and critical bosons, where the mass of the fermion can be used to set the lattice spacing.  The continuum limit is taken by tuning $U$ towards the origin. As far as we know these features of the two dimensional model with $m=0$ were never studied using the Monte Carlo method even at $\mu=0$ where there is no sign problem. The similarity of the model with QCD makes it an interesting toy model for studies at non-zero chemical potential. At a large value of $m$, this was done recently in two space-time dimensions \cite{Li:2016xci}.

\subsection{The Auxiliary Field Representation}

The traditional approach to solve these models is by rewriting the partition function using an auxiliary field formulation so that it can be tackled by the Hybrid Monte Carlo algorithm. More explicitly
\begin{equation}
Z \ =\  \int \ [d\overline{\chi} d\chi]\  e^{-S} \ =\ 
\ \int \ [d\overline{\chi} d\chi] \ \int \ [dA]\  e^{-S_{\rm aux}} 
\label{partitionfunction}
\end{equation}
where in the last step we have introduced a compact auxiliary field $0  \leq A_{x,\nu} < 2\pi$  associated with the bonds of the lattice and the auxiliary field action
\begin{equation}
S_{\rm aux} = \sum_{x,\nu} \frac{N_F}{g^2}\left ( 1-\cos A_{x,\nu} \right ) 
+ \sum_{x,y} \overline{\chi}_x (\tilde{M}_{x,y} + m' \delta_{x,y}) \chi_y,
\label{auxact} 
\end{equation}
is now a Gaussian in the Grassmann fields. The Dirac matrix $\tilde{M}_{x,y}$ is defined as
\begin{equation}
\sum_{\nu} \frac {\eta_{x,\nu}}{2} \big( e^{i A_{x,\nu} + \mu \delta_{\nu,0}}  
\delta_{x+\nu,y} - e^{-i A_{y,\nu} - \mu \delta_{\nu,0} } \delta_{x,y+\nu} \big)
\end{equation}
and the parameters $U$ and $m$ are related to $g$ and $m'$ through the relations
\begin{equation}
U= 0.25 \left( \frac{ \MB{0}{\frac{N_F}{g^2}}} { \MB{1}{\frac{N_F}{g^2}} } \right)^2-0.25,
\quad
m = \left ( \frac{ \MB{0}{\frac{N_F}{g^2}}} { \MB{1}{\frac{N_F}{g^2}} } \right )  m'.
\end{equation}
Here $I_0$ and $I_1$ are the Bessel function and the first modified Bessel function. The sign problem in the auxiliary field representation can be traced to the fact that $\mathrm{Det}(\tilde{M} + m')$ does not have any symmetries and can be complex when $\mu \neq 0$, like in QCD. 

\subsection{The Fermion Bag Representation}

\begin{figure} \center
\includegraphics[width=0.45\linewidth]{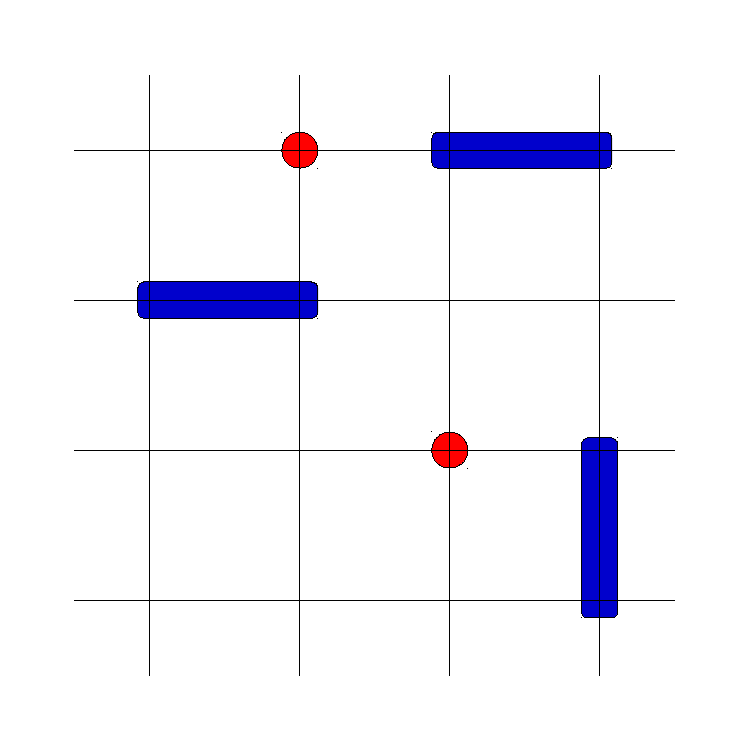}
\caption{ An illustration of a possible configuration of dimers and monomers in a $4\times 4$ block of the lattice. The red circles represent monomer sites and the blue links represent dimers.}
\label{linksandmonomers}
\end{figure}

Can ideas of fermion bags help solve the sign problem present in the auxiliary field approach?  In this approach we do not introduce the usual auxiliary fields, but try to regroup fermion worldlines differently. Unfortunately, this regrouping is not unique and needs some thought. One possible regrouping introduced earlier for $\mu=0$ case is based on introducing a new set of variables, the dimers $d_{x,\nu}$ for nearest neighbor interactions and monomers $n_x$ for the mass terms \cite{Chandrasekharan:2009wc}. This naturally emerges when we expand the Grassmann exponential of the mass and interaction terms:
\begin{align}
Z &= \int d\bar\chi d\chi e^{-\sum_{x,y} \overline{\chi}_x  M_{x,y} \chi_y}  \nonumber \\
&\times \prod_{x} \left ( 1 - m  \bar\chi_{x} \chi_{x} \right ) 
\prod_{x,\nu} \left ( 1- U  \bar\chi_{x} \chi_{x+\nu} \bar\chi_{x+\nu}\chi_x \right ).
\label{grpf}
\end{align}
We then interpret the expression 
\begin{align}
\left ( 1 + m  \bar\chi_{x} \chi_{x} \right ) = \sum_{n_x=0,1} \left ( -m  \bar\chi_{x} \chi_{x} \right ) ^{n_x} .
\end{align}
on each site, as introducing a {\it monomer} field $[n]$ where $n_x=0$ takes values $0$ and $1$. The mass term $(-m \overline{\chi}_x\chi_x)$ is a monomer (single site fermion bag). Similarly the interaction term can be rewritten using a {\it dimer} field $[d]$ such that the interaction term $(-U  \bar\chi_{x} \chi_{x} \bar\chi_{x+\nu}\chi_{x+\nu})$ is the dimer (two site fermion bag). The partition function then becomes the sum over all configurations of $[n]$ and $[d]$. Because of the Grassmann nature of the fermion field, dimers and monomers cannot touch each other. Grassmann fields can be integrated over the monomer and dimer sites first and this does not introduce any sign problems. The remaining Grassmann integral can then be performed on free sites $[f]$ that do not contain monomers or dimers. If we denote the fermion matrix $M$ restricted to the free sites as $W([f])$ we can write the partition function as
\begin{equation}
Z \ =\ \sum_{[d],[n]} m^{N_m} U^{N_d} \mathrm{Det}( W([f]).
\label{fbZ1}
\end{equation}
As an illustration we show a possible configuration of dimers and monomers on a $4\times 4$ block of lattice sites in Fig.~\ref{linksandmonomers}. The monomers are depicted as red circles spanning a single site and the dimers as blue links spanning two sites. The figure depicts a configuration with two free fermion bags that are isolated from each other by the dimers and monomers. Due to this the matrix $W([f])$ is block diagonal with block matrices $W_1[f_1]$ and $W_2[f_2]$ defined within the two independent free bags. The determinant of $W([f])$ is then the product of two determinants $\det(W([f]) = \prod_i \det(W_i([f_i]))$.

When $\mu=0$, since the matrices $W([f])$ are always anti-symmetric, $\mathrm{Det}(W([f]) \geq 0$ and the sign problem is solved. However, in the case of $\mu \neq 0$ this property no longer holds and the determinants can be negative. This may seem surprising since in two space-time dimensions the fermion permutation sign is absent due to the fact that fermions cannot cross each other. In our model fermions have a flavor and they can change flavors while hopping. This is encoded in the staggered phase factors and this leads to a sign problem. Empirically we discovered that this remaining sign problem depends on the boundary conditions. While the sign problem is present with both periodic and anti-periodic boundary conditions, it is absent with open boundary conditions. This also means that on large space-time lattices with $L_X=L_T$ the sign problem essentially disappears, but for  asymmetric lattices it can reemerge. In the most interesting case for our studies, where we fix the spatial lattice size $L_X$, and study very large values of $L_T$ the sign problem can become severe with periodic and anti-periodic boundary conditions.

\subsection{ World Line Representation }

In order to get a better understanding of the origin of the sign problem in our model, we look at the representation of the fermion determinant $\mathrm{Det}(W[f])$ inside free fermion bags as a sum over their world lines. This representation can be found by expanding the determinant back into the Grassmann integral form,
\begin{align}
&\det\left( W([f]) \right ) \label{wl_expanded}\\
&= \prod_{x\in[f]} \left ( \int d\bar\chi_x d\chi_x \right ) e^{-\sum_{x,y \in [f]} \overline{\chi}_x  M_{x,y} \chi_y} \nonumber \\
& =  \prod_{x\in[f]} \left ( \int d\bar\chi_x d\chi_x \right ) \prod_{x,x+\nu \in [f]} \nonumber \\
&\Big( 1 - \frac 12 \eta_{x,\nu}  e^{\mu \delta_{\nu,0} }  \overline{\chi}_x \chi_{x+\nu} + \frac 12 \eta_{x,\nu}^\dagger e^{-\mu \delta_{\nu,0} }  \overline{\chi}_{x+\nu} \chi_{x} \Big). \nonumber
\end{align}
This product can be represented in terms of directed fermion link variables $l_{x,\pm \nu} = 0,\pm 1$, where $+1$ represents the term $\overline{\chi}_x\chi_{x\pm\nu}$ and $-1$ represents the term $\overline{\chi}_{x+\nu}\chi_x$. The determinant is replaced with a sum over all configurations of directed links.

\begin{figure}[b]
\includegraphics[width=0.45\linewidth]{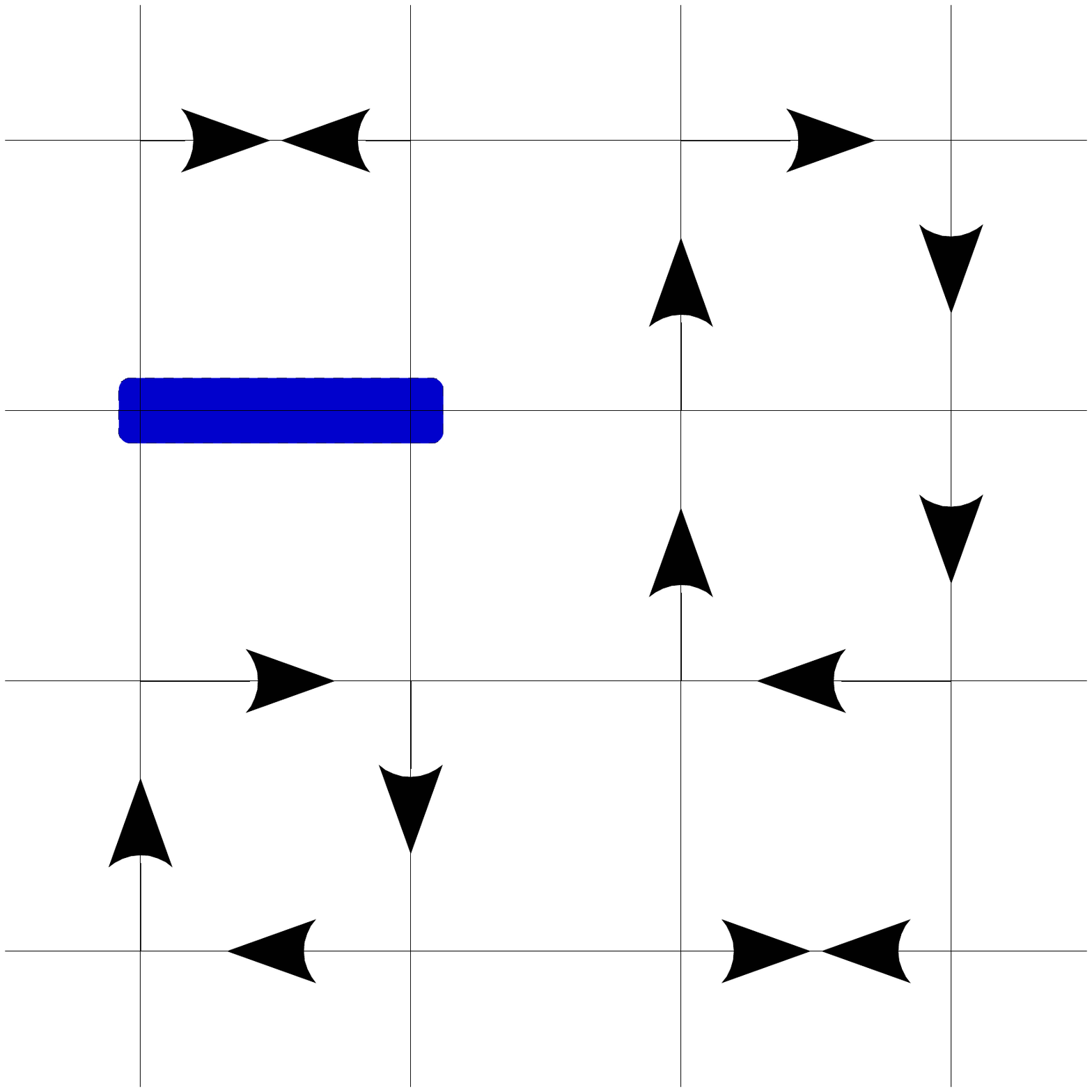} \,\,\,\,
\includegraphics[width=0.45\linewidth]{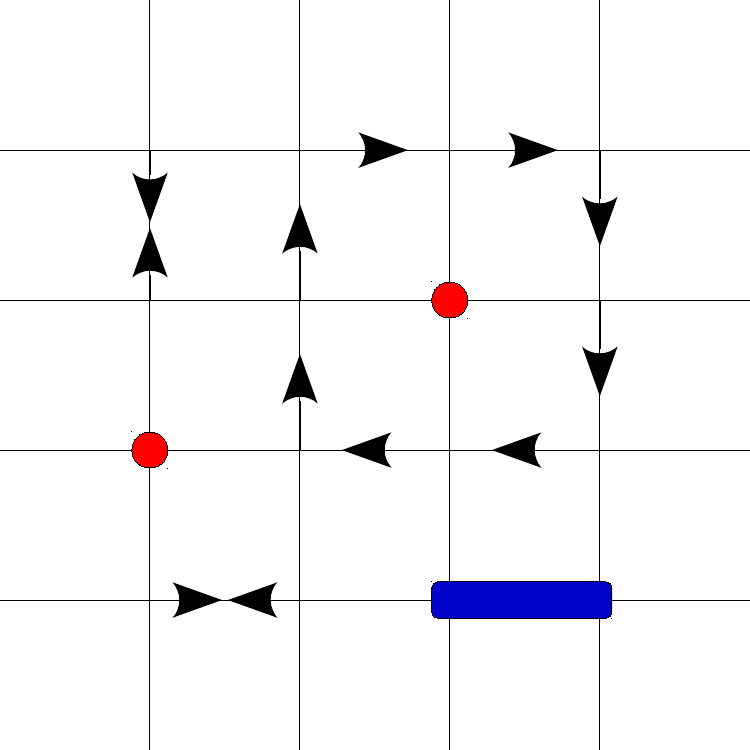}
\caption{ Illustration of two fermion world line configurations with along with dimers and monomers.}
\label{wordlineswithlink}
\end{figure}

Configurations of links only have a nonzero weight when one $\bar \chi$ and one $\chi$ are chosen at each site. Thus each site must have one directed link pointing into it and one pointing out of it. The links will therefore form closed loops.. In Fig.~\ref{wordlineswithlink} we show two valid configurations with the directed links represented as arrows pointing from $\overline{\chi}$ to $\chi$.  The weight of a configuration of fermion world lines is given by the product of the weights in Eq. \ref{wl_expanded} and a factor $-1$ for every closed loop arising from a reordering of $\chi_x$ and $\bar \chi_x$ to match the ordering of the measure.
\begin{align}
&\det\left( W([f]) \right ) \label{wlweight} \\
&= \sum_{[l]} (-1)^{N_{loops}} 
\prod_{x,\alpha} \left ( e^{ - \mu l_{x,\alpha} \delta_{\alpha,0}} \frac {l_{x,\alpha} \eta_{x,\alpha}}2 \right )^{|l_{x,\alpha}|}, \nonumber 
\end{align}
where $N_{loops}$ is the number of closed loops formed by the directed links.
It is easy to verify that there are valid configurations with a negative weight. For example, the configuration on the left in Fig.~\ref{wordlineswithlink} has a positive weight, but the configuration on the right has a negative weight. 

Let us now prove that the sign problem disappears with open boundary conditions in the massless limit because configurations with a negative sign are absent at the worldline level. The weight of a configuration can be written as the product of the weights of the closed loops of fermion links
\begin{align}
&\det\left( W([f],\mu) \right ) \\
&= \sum_{[l]}  \prod_{loop \in l} \left(-\prod_{x,\alpha \in loop}   e^{- \mu l_{x,\alpha} \delta_{\alpha,0}} \frac {l_{x,\alpha} \eta_{x,\alpha}}2 \right ). \nonumber
\end{align}
It is therefore sufficient to show that all loops that can exist in a configuration have positive weight.

\begin{figure} \center
\includegraphics[height=0.3\linewidth]{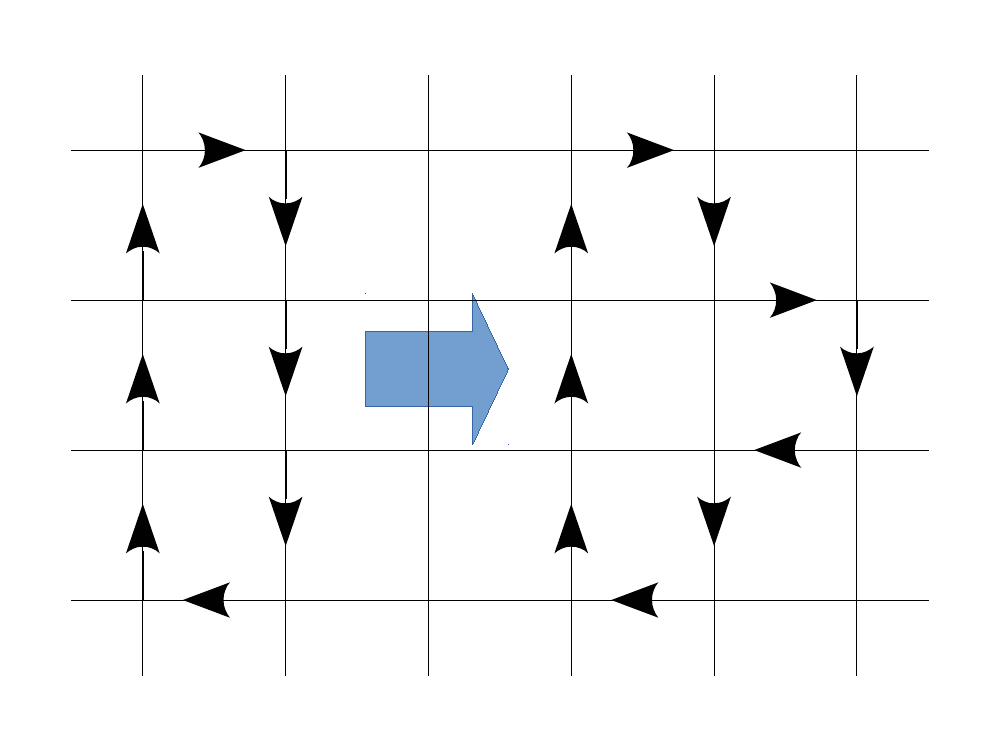} \,\,\,\,
\includegraphics[height=0.3\linewidth]{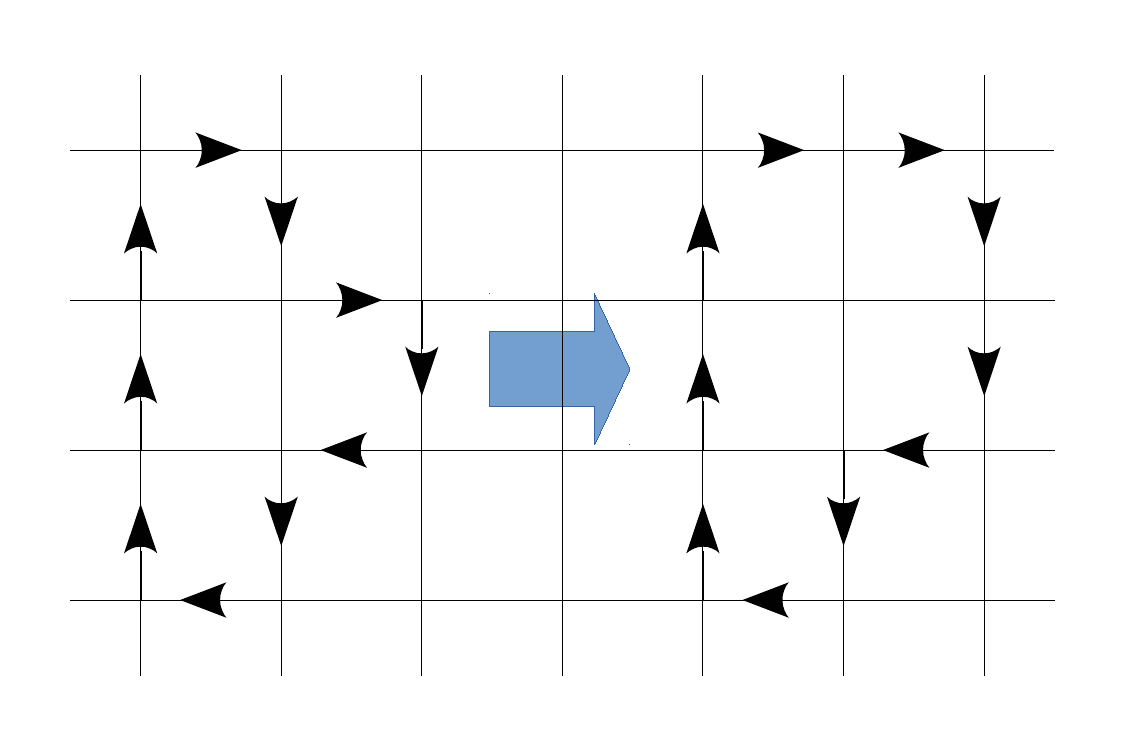}
\caption{ Two deformations that can be used to link any two loops with the same amount of spatial and temporal wrappings. The first replaces a link with a staple and never changes the sign of the loop. The second changes the order of two orthogonal links, changing the sign of the loop. This deformation also changes the number of enclosed sites by one.}
\label{deformations}
\end{figure}

Consider first a loop that does not wrap around the volume. Note that by starting from the trivial loop that visits two neighboring sites, we can construct any non-wrapping loop using the two deformations depicted in Fig.~\ref{deformations}. The first deformation replaces a link with a staple and does not introduce any new sites inside. This deformation also does not change the sign of the loop. The second deformation inverts a corner and introduces a single site inside the loop. This does change the sign of the loop. Thus, any non-wrapping loop can be negative only if it encloses an odd number of sites. But in the massless limit a configuration with such a loop will have zero weight, since monomers are not allowed, dimers always take away two sites and all free fermion loops will touch an even number of sites. Given that the trivial loop has a positive sign, any allowed non-wrapping loop will have a positive sign. Similarly, any loop enclosing an odd number of sites has a negative sign.

\begin{figure} 
\includegraphics[height=0.4\linewidth]{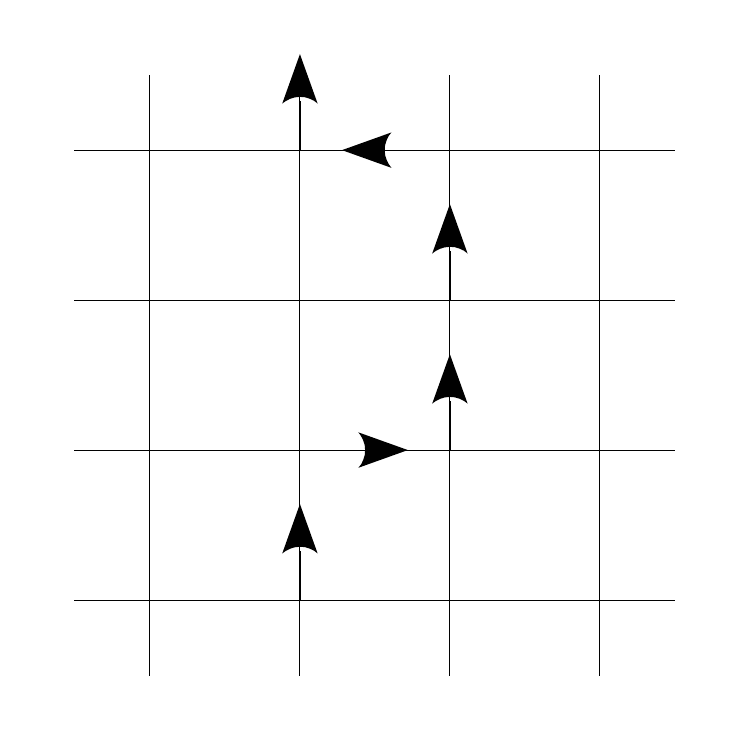} \,\,\,\,
\includegraphics[height=0.4\linewidth]{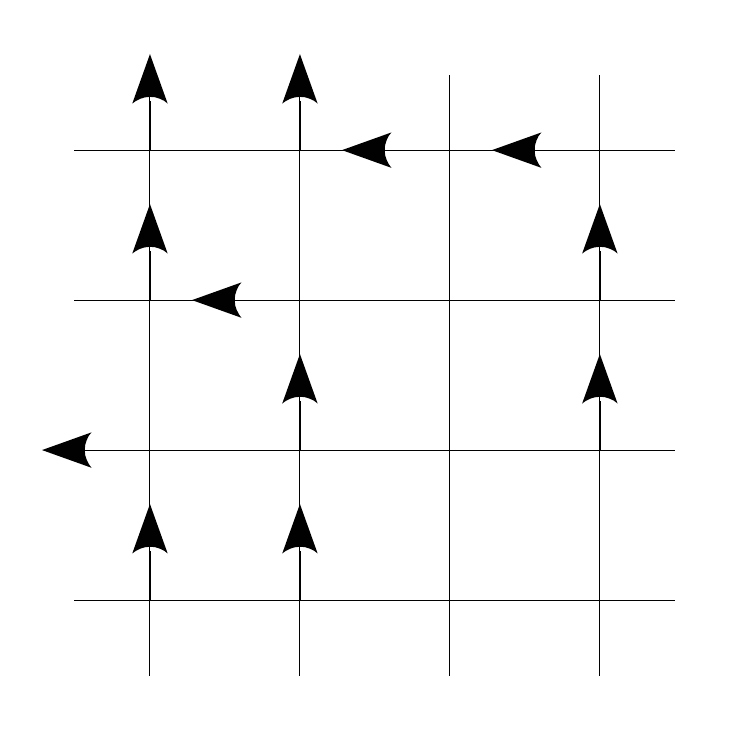}
\caption{The figure on the left shows a negative sign fermion loop that wraps along the temporal direction. Such a loop is generated when an odd number of sites cross the fermion world line as it is obtained through a series of deformations starting from a straight temporal loop. The figure on the right shows a loop with negative sign when the spatial boundary condition is (anti)symmetric. }
\label{negativeloops}
\end{figure}

Thus, any negative sign fermion loops in the massless theory must arise through loops that wrap around the temporal direction. Note that with open boundary conditions spatial winding is also forbidden. We can again construct any temporal wrapping loop by starting from a loop that goes straight in time without hops and deforming it using the two deformations discussed above. This time a negative sign in the loop can be introduced if an odd number of sites cross the fermion line during this deformation. An example of such a negative signed loop is shown in Fig.~\ref{negativeloops} on the left. Such a loop will be allowed if the left and right sides of the loop are connected through the boundary. However, with open boundary conditions such temporal loops will create regions on the left and right with an odd number of sites. This is forbidden in the massless limit for the same reasons as outlined above for non-wrapping loops.

With periodic and anti-periodic boundary conditions we can have other more complicated loops as shown in  Fig.~\ref{negativeloops} on the right. Thus, the sign problem can be completely eliminated by open boundary conditions in the spatial direction. This feature of the world line formulation is well known and specific to two dimensional models \cite{Evertz:2000rk,Gattringer:2007em,Wolff:2007ip}. In higher dimensions the argument for the positivity of all fermion worldline configurations fails and significant cancellations between world line configurations will be necessary for alleviating the sign problem. The fermion bag approach can be helpful in this regard \cite{Li:2016xci}.

\begin{figure} 
\includegraphics[height=0.25\linewidth]{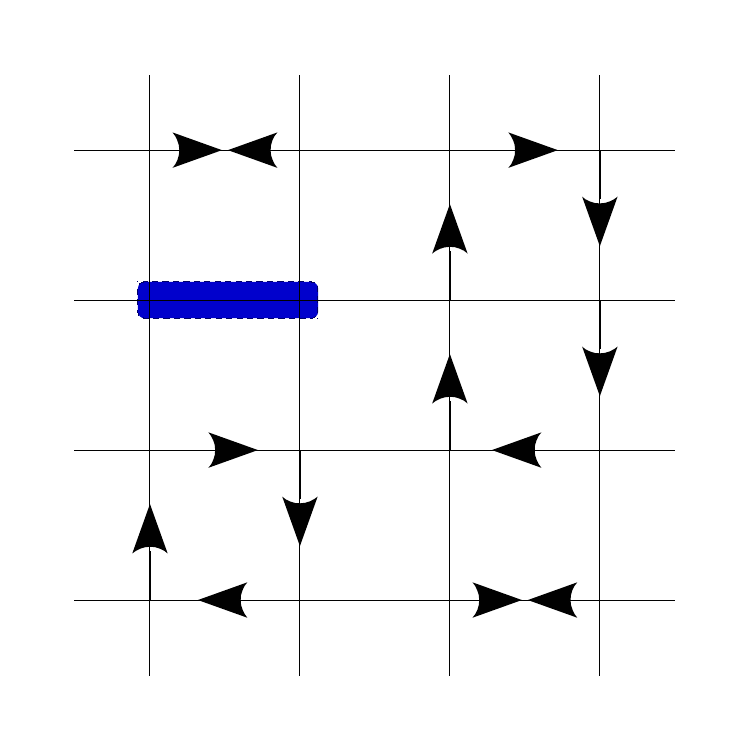}
\includegraphics[height=0.25\linewidth]{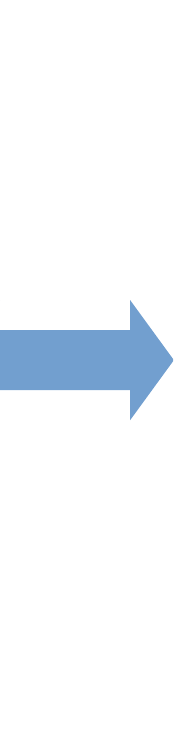}
\includegraphics[height=0.25\linewidth]{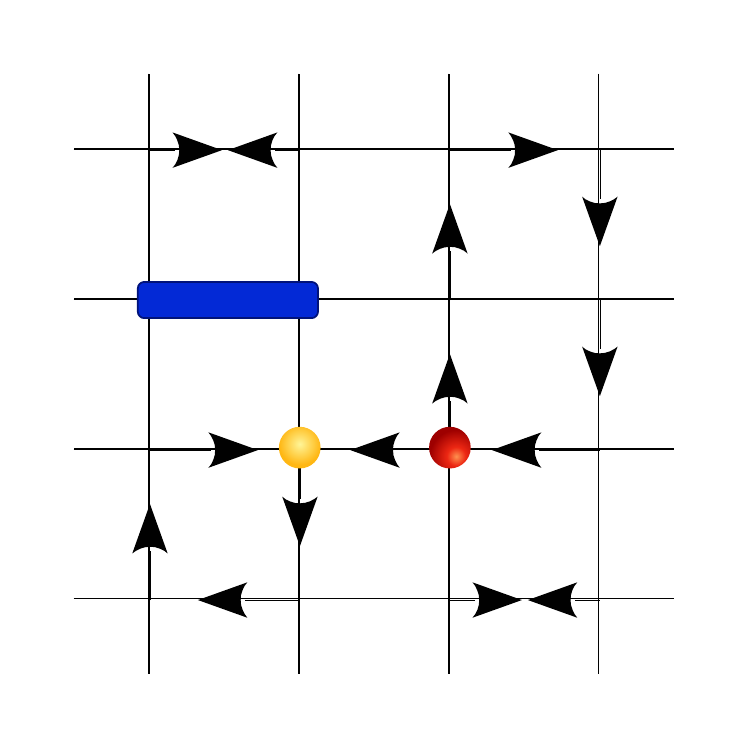}
\includegraphics[height=0.25\linewidth]{figs/fig6m.pdf}
\includegraphics[height=0.25\linewidth]{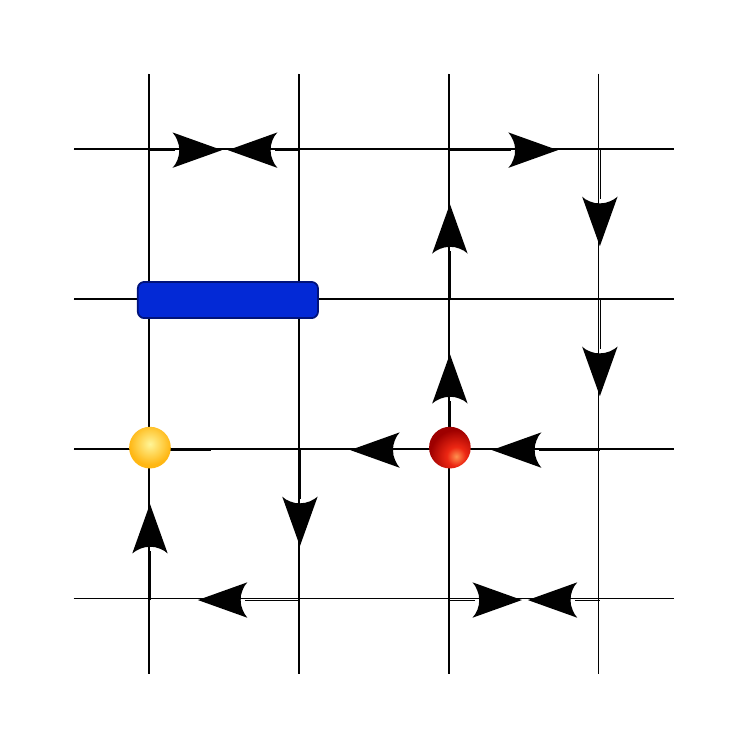}\\
\includegraphics[height=0.25\linewidth]{figs/fig6m.pdf}
\includegraphics[height=0.25\linewidth]{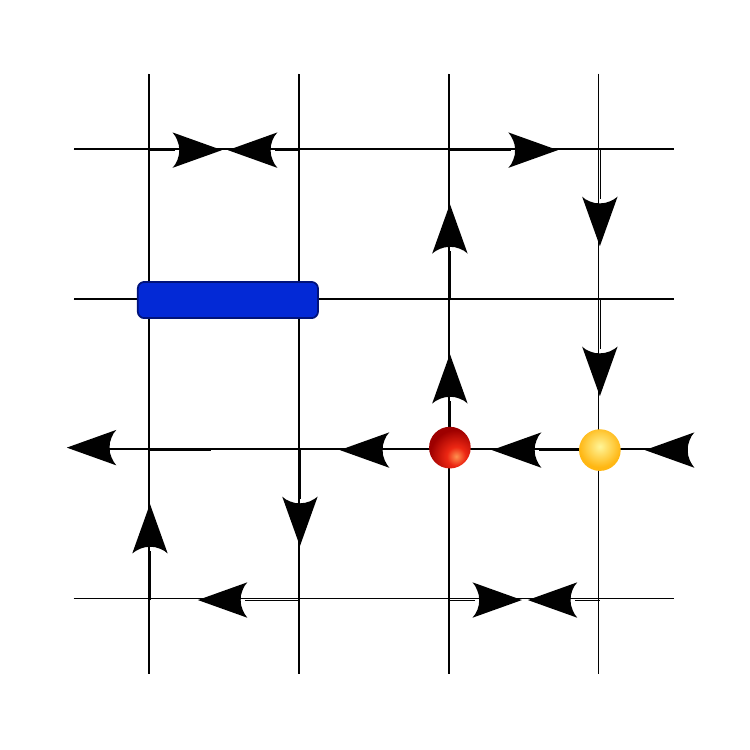}
\includegraphics[height=0.25\linewidth]{figs/fig6m.pdf}
\includegraphics[height=0.25\linewidth]{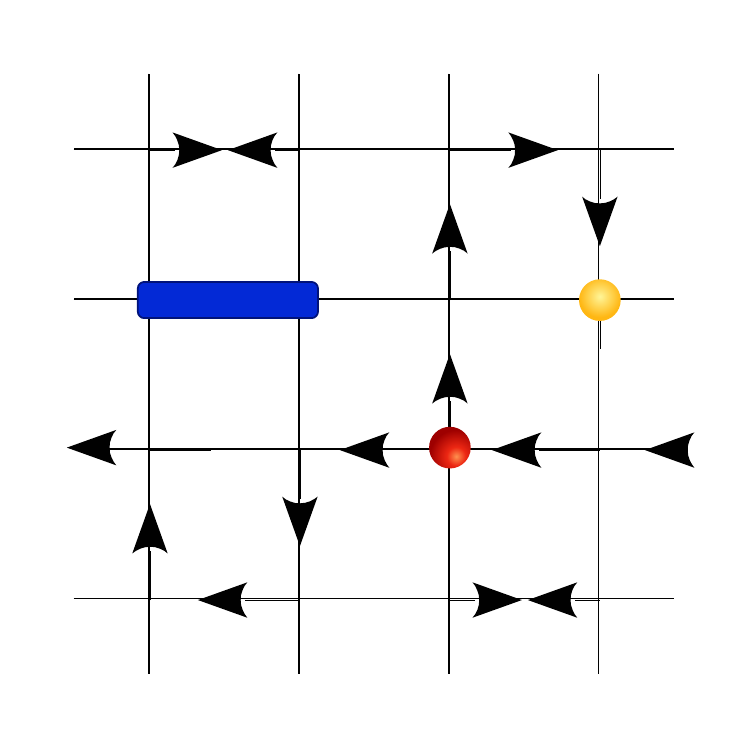}
\\\vspace{0.2cm}
\includegraphics[height=0.25\linewidth]{figs/fig6m.pdf}
\includegraphics[height=0.25\linewidth]{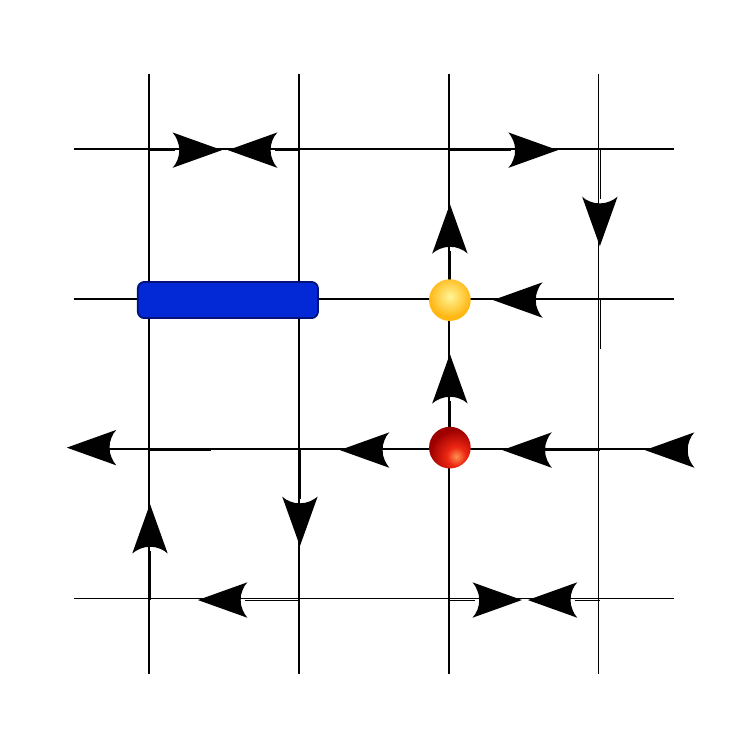}
\includegraphics[height=0.25\linewidth]{figs/fig6m.pdf}
\includegraphics[height=0.25\linewidth]{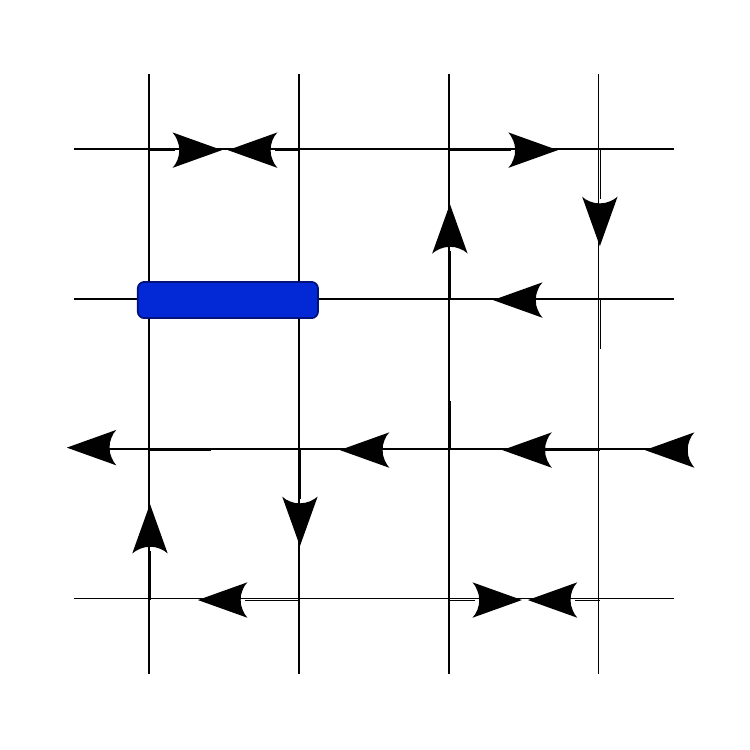}
\caption{An illustration of the worm update used in the world line formalism. The solid dots represent the head and the tail of the worm where the configuration has defects. At the end the defects disappear and a new allowed configuration is generated. }
\label{wlupdates}
\end{figure}

\section{Monte Carlo Updates}
\label{mcmethods}

\begin{figure} 
\includegraphics[height=0.25\linewidth]{figs/fig6a.pdf}
\includegraphics[height=0.25\linewidth]{figs/fig6m.pdf}
\includegraphics[height=0.25\linewidth]{{figs/fig6b}.pdf}
\includegraphics[height=0.25\linewidth]{figs/fig6m.pdf}
\includegraphics[height=0.25\linewidth]{figs/fig6c.pdf}\\
\includegraphics[height=0.25\linewidth]{figs/fig6m.pdf}
\includegraphics[height=0.25\linewidth]{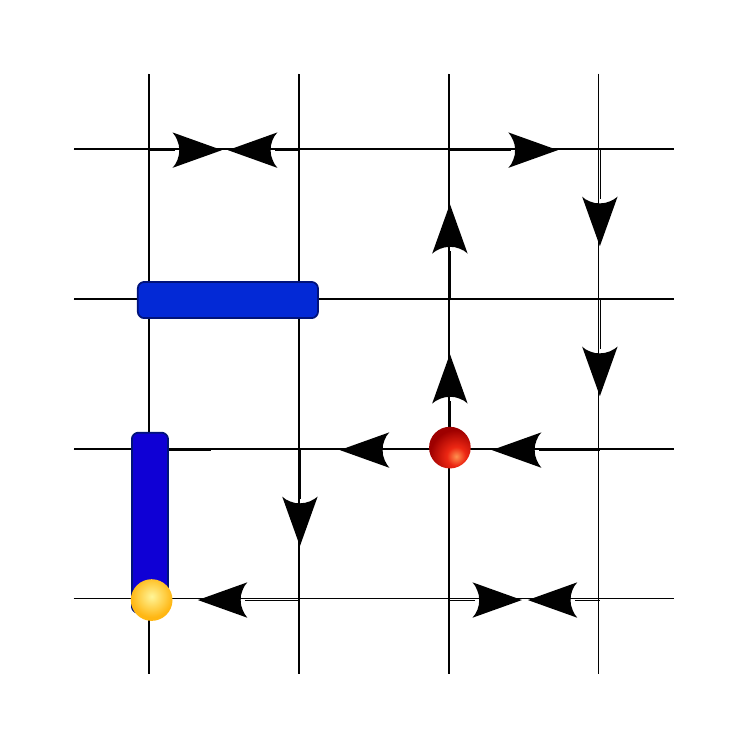}
\includegraphics[height=0.25\linewidth]{figs/fig6m.pdf}
\includegraphics[height=0.25\linewidth]{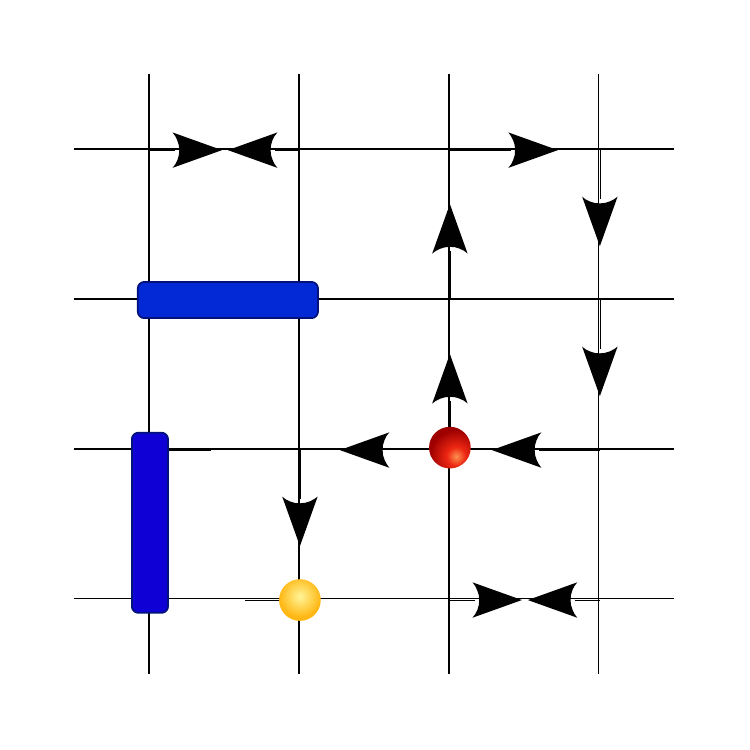}
\\\vspace{0.2cm}
\includegraphics[height=0.25\linewidth]{figs/fig6m.pdf}
\includegraphics[height=0.25\linewidth]{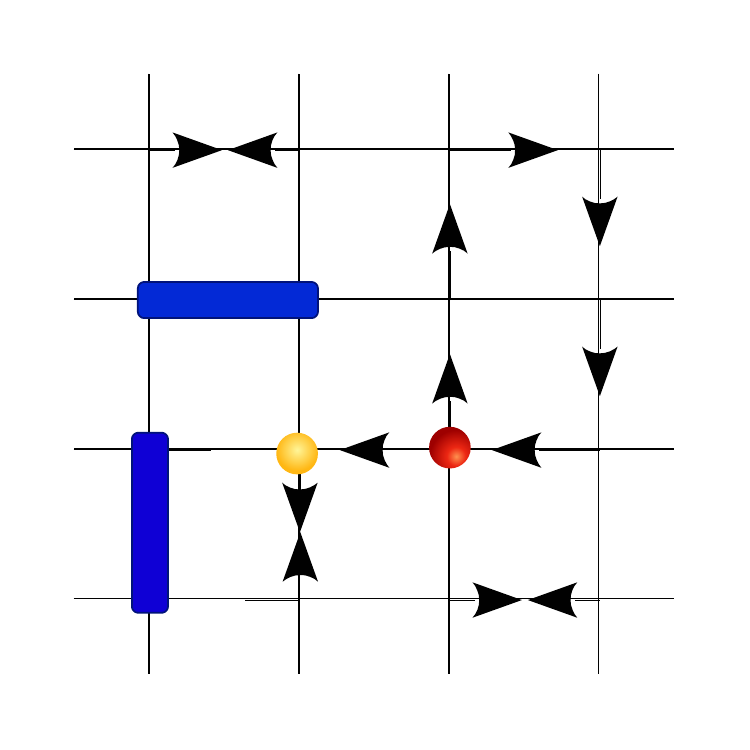}
\includegraphics[height=0.25\linewidth]{figs/fig6m.pdf}
\includegraphics[height=0.25\linewidth]{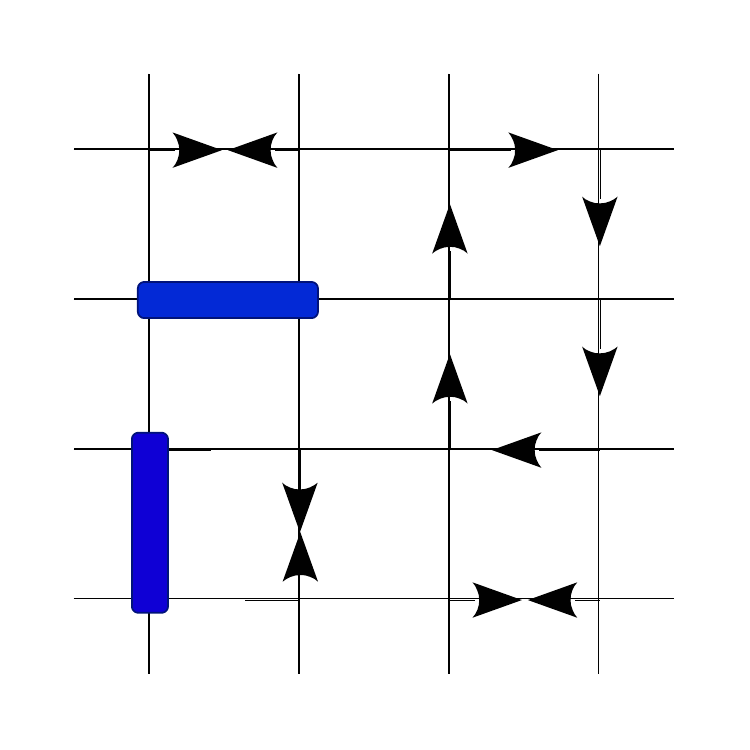}
\caption{An illustration of the worm update that changes the dimer number. }
\label{wldimerud}
\end{figure}

Monte Carlo methods for updating both the worldline representation and the fermion bag representations are by now well developed \cite{PhysRevLett.87.160601,PhysRevE.66.046701,PhysRevE.74.036701,PhysRevLett.108.140404,PhysRevD.88.021701,PhysRevD.93.081701}. We use a worm algorithm to update the fermion lines and dimers, using updates like the one illustrated in Fig.~\ref{wlupdates}. To begin an update, we suggest randomly changing some fermion link $l_{x,\nu}$. The update is then accepted with the absolute value of the weight given in Eq. (\ref{wlweight}). If the link is changed two defects are generated in the lattice configuration which are allowed. The defects are the head and tail of the worm. The head of the worm then propagates by updating the neighboring links. When the head returns to its tail, the worm closes, the defects disappear and the update is complete. The various steps of how the defect propagates are shown in shown Fig.~\ref{wlupdates}. The configuration of dimers may also be updated during the worm update. When this is done we have to use the weights of including or removing a dimer. Fig.~\ref{wldimerud} shows the steps for an update that changes the dimer number.

In contrast to the worm algorithm, we sample fermion bag configurations using a local Monte Carlo update that involves adding or removing dimers or pairs of monomers. Each proposal is accepted with the probability 
\begin{align}
P_{acc} = \frac{m^{N'_m} U^{N'_d} \det\left( W([f'],\mu) \right ) }{m^{N_m} U^{N_d} \det\left( W([f],\mu) \right )},
\end{align}
where the new configuration is denoted with primed variables. The fact that one has to use ratios of fermion determinants that are non-local helps in reducing autocorrelation times. We also can update large regions of space-time by using a background field method used recently in \cite{PhysRevD.93.081701}. The sampling is made more efficient with a move that switches the places of a monomer and a dimer if the two are on neighboring sites. Since the weights of the two configurations are the same this update is very quick.

\section{Numerical results} 
\label{results}

In this work we compute three observables in order to understand the physics of our model. The first is the chiral condensate susceptibility $\chi$, defined by the relation
\begin{align}
\chi = \frac{U}{V} \sum_{x,y} \ev{\bar\psi_x\psi_x \bar\psi_y\psi_y}\label{susceptibility}.
\end{align}
We can use it to understand the physics of bosonic excitations in our model. We also compute the chiral charge winding number susceptibility, defined by the relation
\begin{align}
\ev{Q_\chi^2} &= \frac U V \sum_{x \in S,y \in S'} \ev{J_{\alpha,x}^\chi J_{\alpha,y}^\chi},\\
J_{\alpha,x}^\chi &= \frac{\epsilon_x \eta_{x,\alpha}} 2 \left [ e^{\delta_{\alpha,0}\mu} \bar\psi_x \psi_{x+\alpha} - e^{-\delta_{\alpha,0}\mu}\bar\psi_{x+\alpha} \psi \right ]
\end{align}
where $S$ and $S'$ are surfaces orthogonal to the direction $\alpha$. In the thermodynamic limit, the winding number susceptibility helps us understand the status of chiral symmetry as we explain below. In the world line representation the chiral charge can be defined by the relation
\begin{align}
q_{x_\alpha}^\chi &= \epsilon_x ( l_{x,\alpha} + l_{x+\alpha,-\alpha} + 2d_{x,\alpha} ),
\end{align}
which means the susceptibility is simply
\begin{align}
\ev{Q_\chi^2} =  \ev{\left ( \sum_{x\in\alpha} q_{x_\alpha}^\chi \right )^2 }
\end{align}
since the chiral charge is conserved on each configuration. Finally we measure the average fermion number using the relation $\langle N_f \rangle = \langle \sum_{x\in S} J_{0,x} \rangle$ where the fermion number current is given by
\begin{align}
J_{\alpha,x} &= \frac{ \eta_{x,\alpha}} 2 \left [ e^{\delta_{\alpha,0}\mu} \bar\psi_x \psi_{x+\alpha} - e^{-\delta_{\alpha,0}\mu}\bar\psi_{x+\alpha} \psi_x \right ],
\end{align}
and $S$ is a surface perpendicular to $\hat t$. In the worldline representation again the fermion number is straight forward to calculate and is given by
\begin{align}
\ev{N_f} = \sum_{x\in S} \ev{ l_{x,\hat t} - l_{x+\hat t,-\hat t} }.
\end{align}
In our definition the fermion number is normalized to count both the Dirac and flavor degrees of freedom from a continuum limit perspective. 

\begin{table}
\center
\begin{tabular}{|c|c|c|c|}
\hline
$U$ & $2-\eta$ & $\ev{Q_\chi^2}$ & $m_f$ \\ 
\hline
0    & 0        & 0.25     & 0          \\
0.1  & 0.90(1)  & 0.499(7) & 0.0098(4)  \\
0.2  & 1.201(4) & 0.61(1)  & 0.081(1)   \\
0.3  & 1.303(4) & 0.780(8) & 0.183(1)   \\
0.4  & 1.371(7) & 0.895(4) & 0.290(1)   \\
0.5  & 1.393(3) & 0.972(3) & 0.395(3)   \\
0.6  & 1.423(4) & 1.024(3) & 0.491(1)   \\
1.0  & 1.467(4) & 1.128(2) & 0.793(1)   \\
$\infty$ & 1.5  & 1.208(8) & $\infty$  \\
\hline
\end{tabular}
\caption{ The scaling dimension $\nu$, chiral charge susceptibility $\ev{Q_\chi^2}$ and the fermion mass measured on a square lattice. }
\label{table_large_volume}
\end{table}

\begin{figure}[b]
\includegraphics[width=0.8\linewidth]{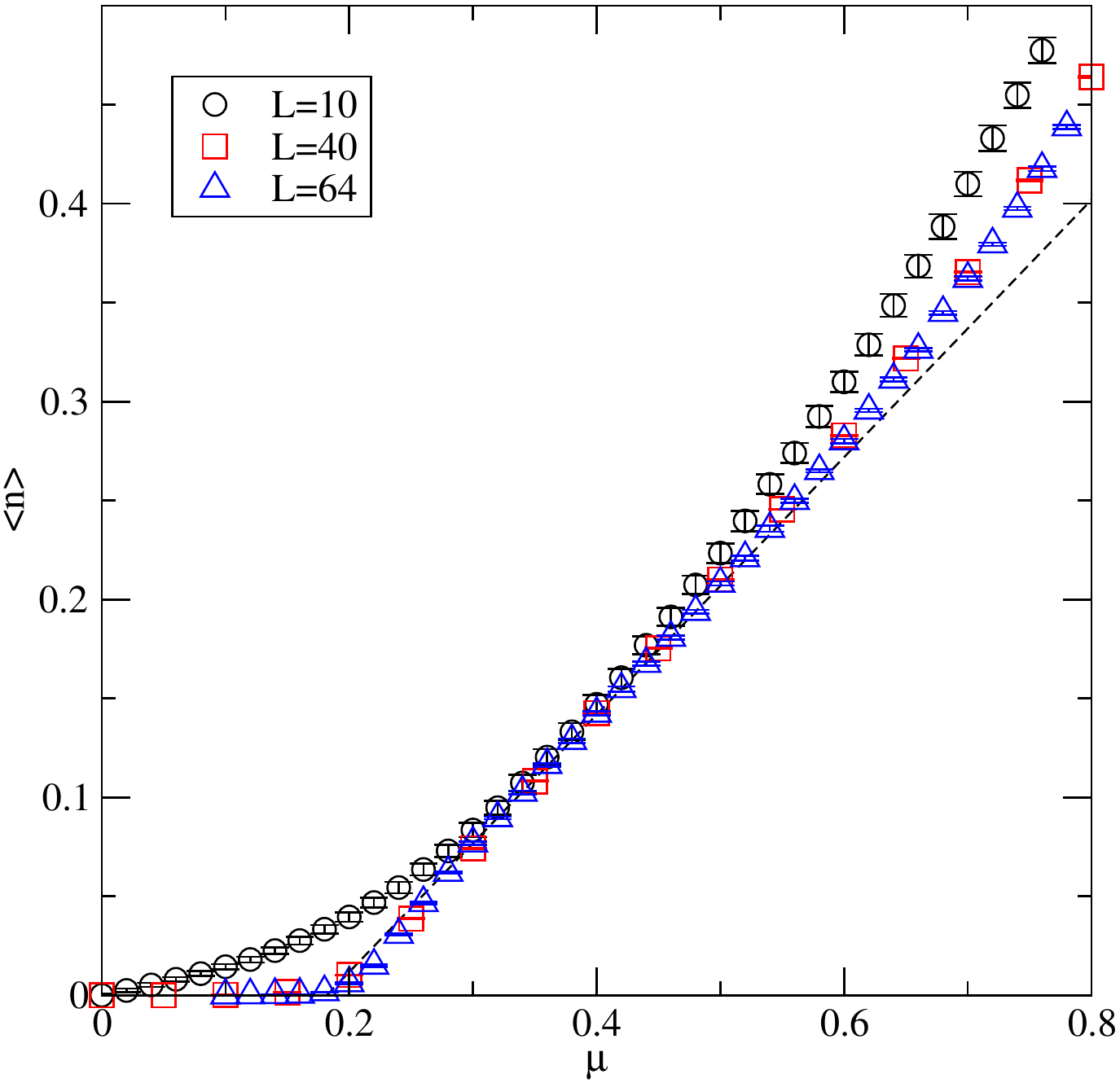}
\caption{ The fermion number density $\ev{n}$ as a function of the chemical potential on a square lattice with open boundary conditions. The dashed line shows a fit to the linear region at $L=64$.}
\label{open_square}
\end{figure}

\begin{table}[htb]
\begin{tabular}{|c|c||c|c||c|c|}
\hline
$ \mu $ & $\langle n \rangle $  & $ \mu $ & $\langle n \rangle $  & $ \mu $ & $\langle n \rangle $  \\
\hline
\multicolumn{6}{|c|}{L=10}\\
\hline
   0.16 &           $0.0276(2)$ &    0.32 &           $0.0948(4)$ &    0.52 &           $0.2397(5)$ \\
   0.20 &           $0.0396(2)$ &    0.36 &           $0.1204(4)$ &    0.54 &           $0.2582(5)$ \\
   0.24 &           $0.0545(3)$ &     0.40 &           $0.1472(5)$ &    0.56 &          $0.2740(5)$ \\
   0.28 &           $0.0729(3)$ &    0.48 &           $0.2074(5)$ &    0.58 &           $0.2926(5)$ \\
\hline
\multicolumn{6}{|c|}{L=40}\\
\hline
   0.15 &          $0.0016(1)$ &     0.30 &           $0.0742(4)$ &    0.45 &   $0.1753(4)$ \\
   0.20 &          $0.0103(2)$ &    0.35 &           $0.1082(4)$ &     0.50 &   $0.2100(5)$ \\
   0.25 &          $0.0387(4)$ &     0.40 &          $0.1425(4)$ &    0.55 &    $0.2456(4)$ \\
\hline
\multicolumn{6}{|c|}{L=64}\\
\hline
   0.16 &          $0.0004(1)$ &    0.32 &         $0.089(1)$ &    0.52 &            $0.221(1)$ \\
   0.20 &          $0.0062(4)$ &    0.36 &         $0.116(1)$ &    0.54 &            $0.236(1)$ \\
   0.24 &          $0.0309(6)$ &    0.40 &         $0.142(1)$ &    0.56 &            $0.250(1)$ \\
   0.28 &          $0.0618(6)$ &    0.48 &         $0.194(1)$ &    0.58 &            $0.264(1)$ \\
\hline
\end{tabular}
\caption{\label{table_os}  Selected values of $\langle n \rangle$ plotted in Fig~\ref{open_square}.}
\end{table}

\begin{figure} \center
\includegraphics[width=0.7\linewidth]{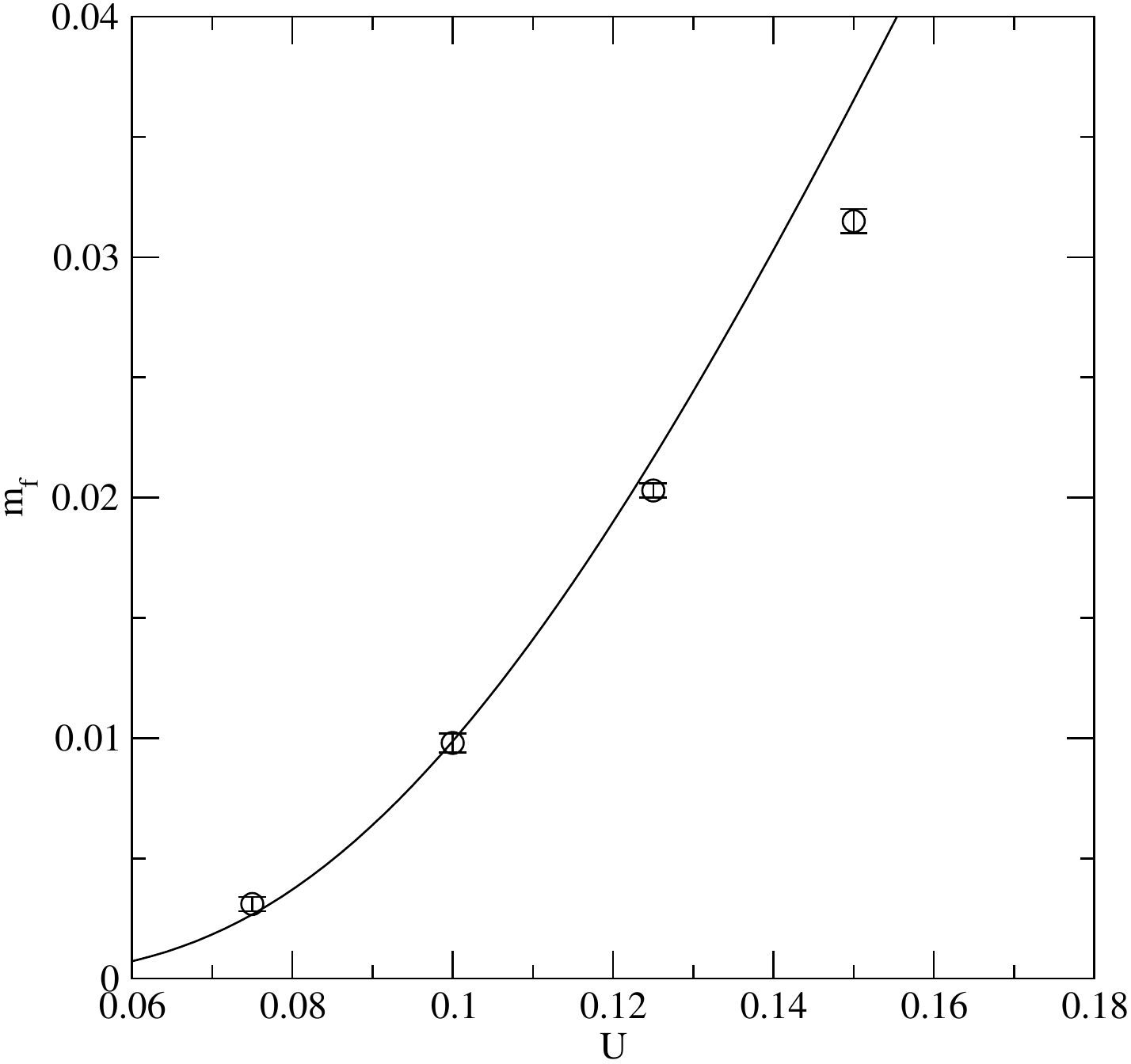}
\caption{Plot of the fermion mass as a function of $U$ for small values. We observe qualitatively the exponential scaling expected. The solid line is the one loop $\beta$ function.}
\label{mf_fit_beta}
\end{figure}

\begin{figure}[t]
\includegraphics[width=0.45\linewidth]{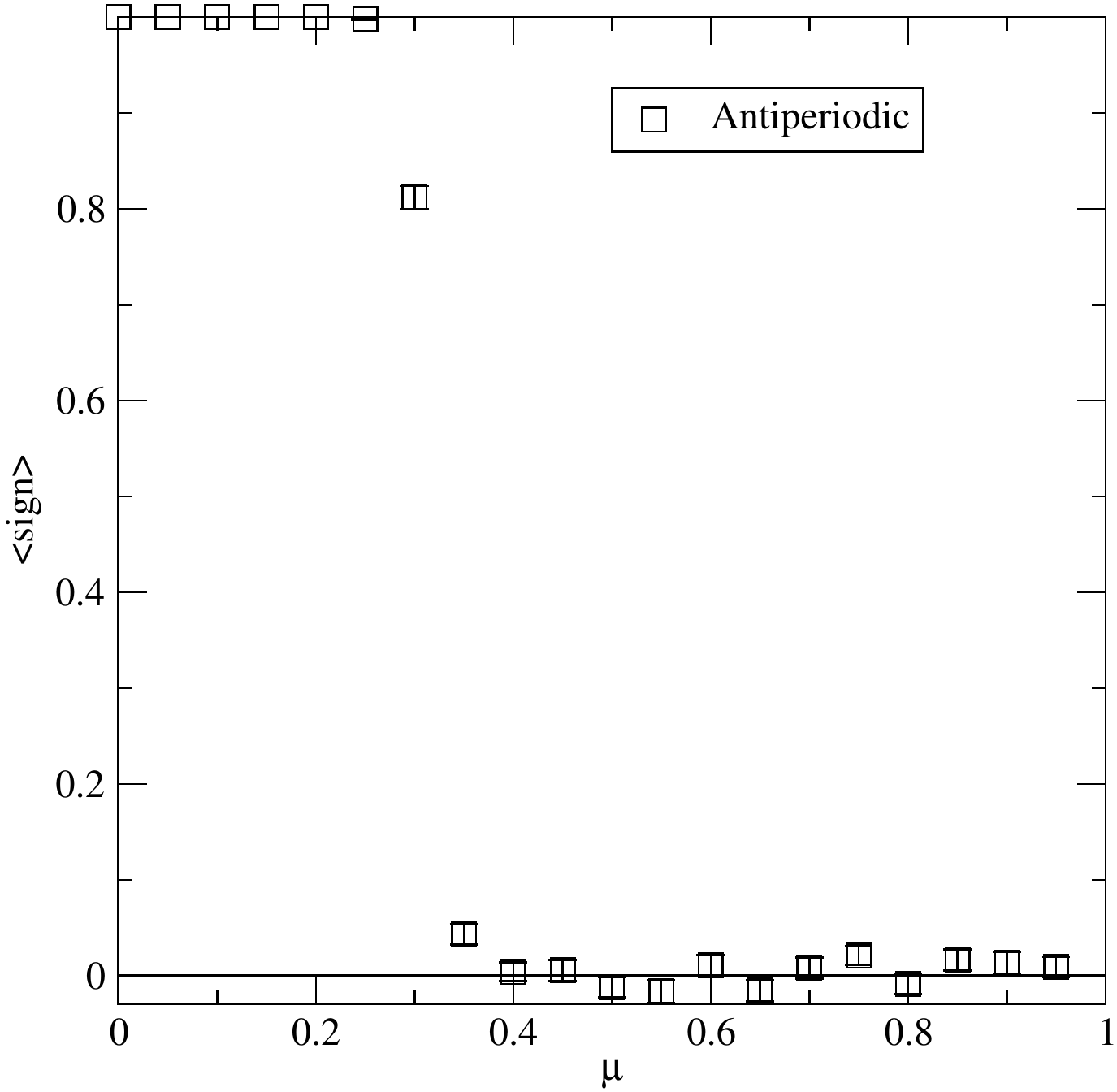}
\includegraphics[width=0.45\linewidth]{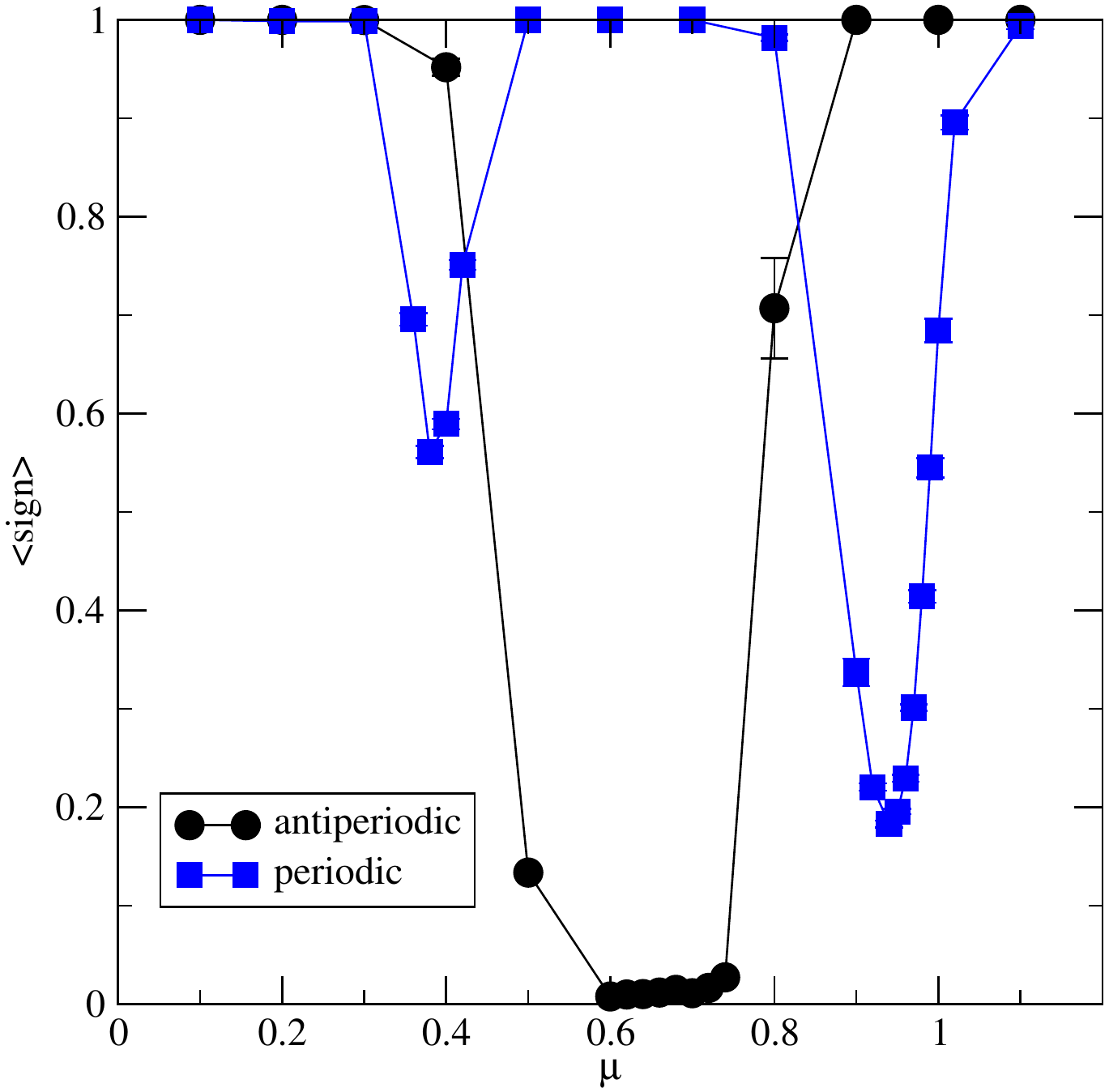}
\caption{ The average sign of $\det(W)$ at $U=0.3$ as a function of the chemical potential with $L_T=48$ and $L_X=6$ in the auxiliary field representation(left) and the fermion bag representation(right). }
\label{sign-boundaries}
\end{figure}

Using these observables we first focus on the physics of our model at $\mu=0$ in order to bring out the similarities to QCD.  As we mentioned earlier, unlike in QCD the $U(1)$ chiral symmetry of the model cannot break in two dimensions. However, the lightest boson in the model is critical (i.e., it is massless but is not a Goldstone boson). Hence when $L_X=L_T=L$ we expect the chiral condensate susceptibility to scale as $\chi \sim L^{2-\eta}$ for large values of $L$. The exponent $\eta$  depends on $U$ like in the usual critical phase of the two dimensional $XY$ model. At infinite $U$ since the Thirring model becomes a closed packed dimer model and we expect $\eta=0.5$ \cite{Cecile:2008nb}. When $U=0$ the susceptibility diverges logarithmically with $L$ and hence $\eta=2$. Our results reproduce this and show how the exponent changes continuously between these two limits. In table \ref{table_large_volume} we give the values of $2-\eta$ obtained at various values of $U$.

\begin{figure*}[t]
\includegraphics[height=0.25\linewidth]{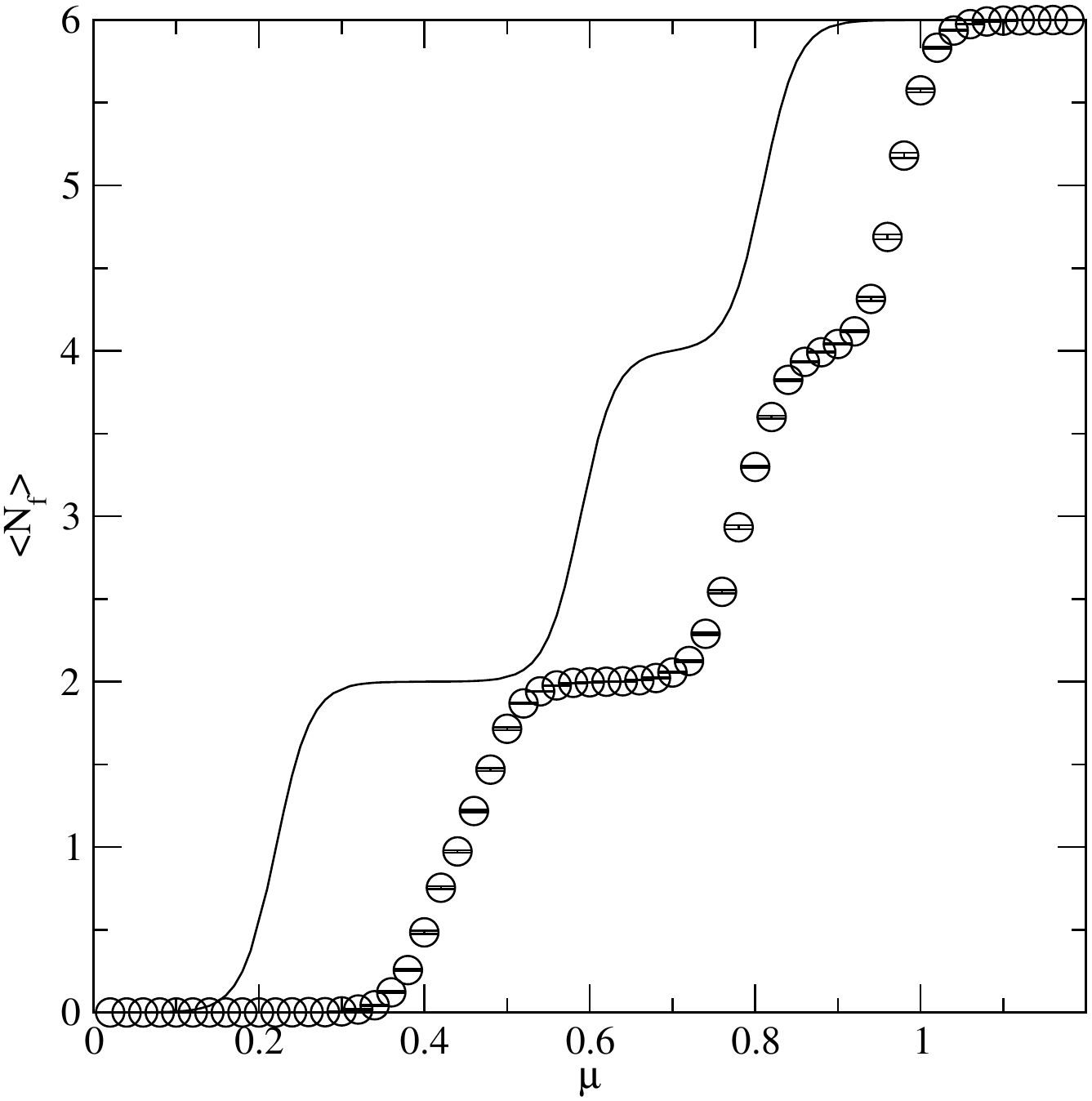}
\includegraphics[height=0.25\linewidth]{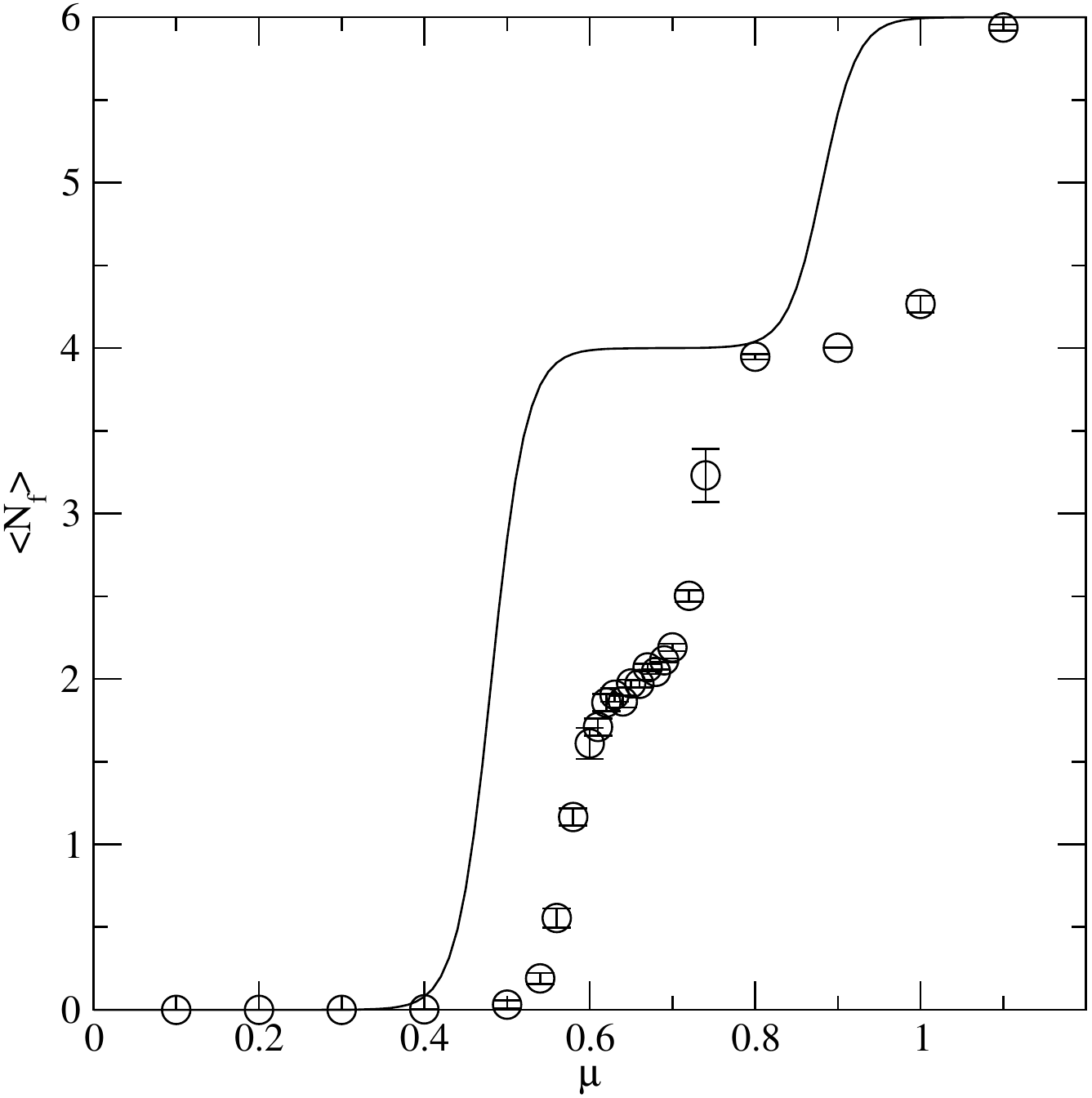}
\includegraphics[height=0.25\linewidth]{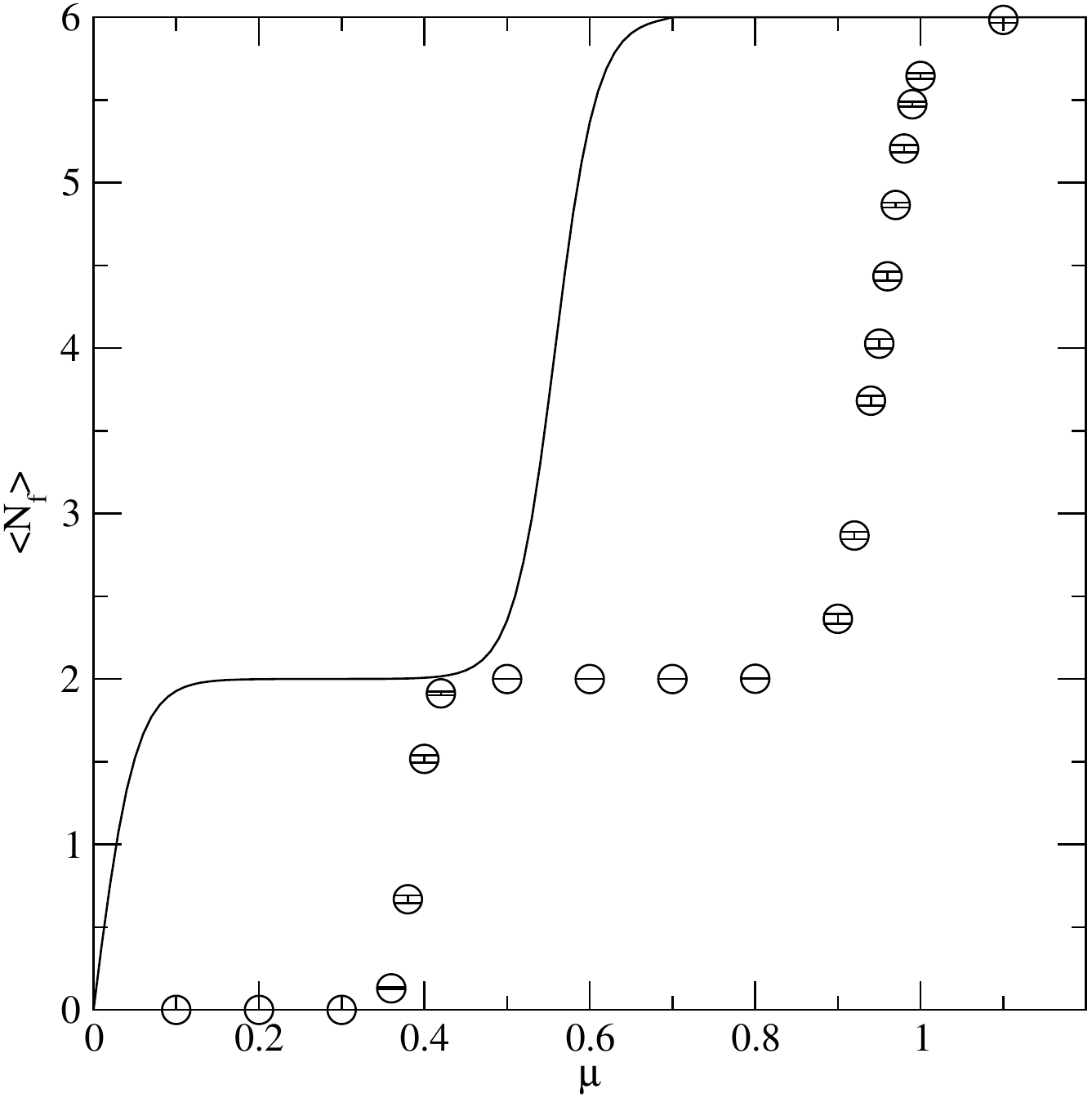}
\caption{ The fermion number $\ev{N_f}$ at $U=0.3$ as a function of the chemical potential with open, antiperiodic and periodic boundary conditions respectively for $ L_{X}=6, L_{T}=48 $. The solid line shows the value at $U=0$.}
\label{nf-boundaries}
\end{figure*}

In a chirally symmetric theory with massive excitations the chiral charge winding number susceptibility $\langle Q_\chi^2\rangle$ is expected to vanish because the chiral charge cannot wind across the spatial boundaries. However, when the phase is critical like in our model it is expected to be go to a constant in the thermodynamic limit. Our results are consistent with this expectation. The values we measured for $\langle Q_\chi^2\rangle$ at $L=256$ are given in table \ref{table_large_volume}. These values are found using open boundary conditions Further we find that $\langle Q_\chi^2\rangle=0.25$ at $U=0$, and grows monotonically to the value of roughly $1.2$ at $U=\infty$. All this is consistent with the fact that the bosonic sector of our theory is critical.

In contrast to the bosons, fermions are massive for all values of $U > 0$. We compute the fermion mass $m_f$ as a function of $U$ using large square lattices $(L_X=L_T=L)$ as follows. In the thermodynamic limit we expect the average fermion density $\langle n\rangle = \langle N_f\rangle/L_X$ to be zero when $\mu \leq m_f$ and rise linearly according to the relation
\begin{align}
\ev{n} = c\left( \mu-m_f \right).
\end{align}
for $\mu \geq m_f$. This behavior should also be an excellent approximation for sufficiently large lattices.  To demonstrate this we show our results for $\langle n\rangle$ at $U=0.3$ with open boundary conditions in Fig.~\ref{open_square}. Selected values of this data are also tabulated in table \ref{table_os} as a benchmark for future calculations. As we can see for $L=10$ the curve does not show the expected non-analyticity, but for $L=40$ and $L=64$ the curves show it clearly. We can fit our data to the linear form which is shown as the dashed line in the fit. In table \ref{table_large_volume} we report the value of $m_f$ found using this method for several values of $U$. We used lattices of  size of $L=64$, except for $U=0.1$, where the lattice size used was $L=128$.

The dynamical generation of fermion mass is an interesting feature of our model. While similar to the phenomenon of chiral symmetry breaking in QCD, the actual dynamical breaking of continuous symmetries is forbidden in two dimensional models. Nevertheless a fermion mass can be generated and a massless boson with critical correlations can arise \cite{Witten:1978qu}. Finally we note that four-fermion couplings are expected to be marginal in two dimensions and in our case it also happens to be marginally relevant (i.e., asymptotically free).  Thus, at small $U$ the fermion mass $m_f$ is expected to vanish according to the relation
\begin{align}
m_f \approx C \ \exp\Big(\frac{-2\pi}{b_0U}\Big),
\end{align}
where $b_0=16$ is the one-loop coefficient of the $\beta$ function. Figure \ref{mf_fit_beta} shows the fermion mass values and compares it against the expected behavior. For purposes of illustration we use $C=0.49$. With  these small masses lattice volumes up to $V=1024\times1024$ were necessary. It is well known that such asymptotic scaling fits don't work very well unless very large lattices are used \cite{PhysRevLett.80.1742}. Here we just use it to illustrate that the the fermion mass qualitatively does become exponentially small as $U$ becomes small.

\begin{figure*}[t]
\includegraphics[height=0.25\linewidth]{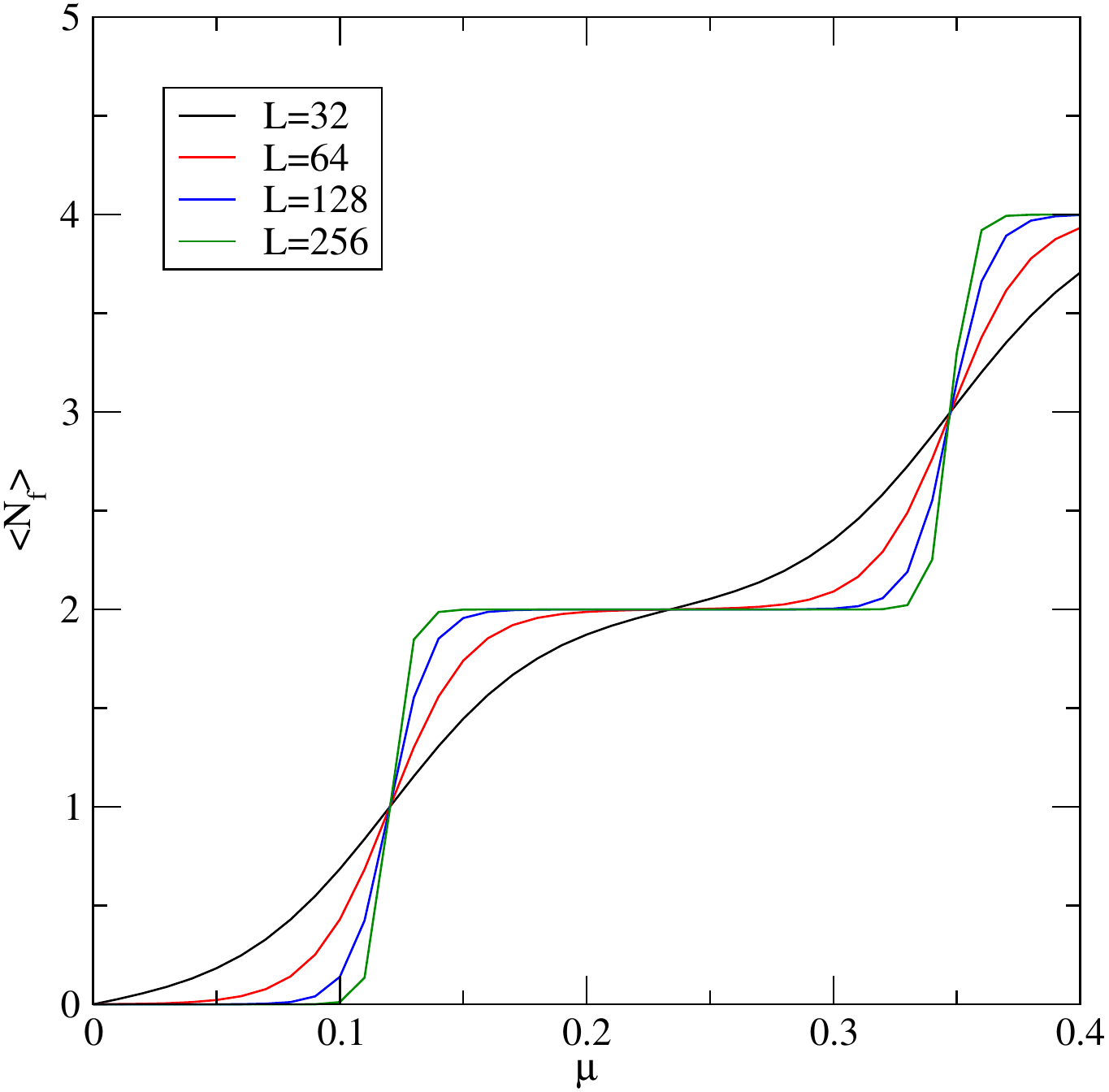}
\includegraphics[height=0.25\linewidth]{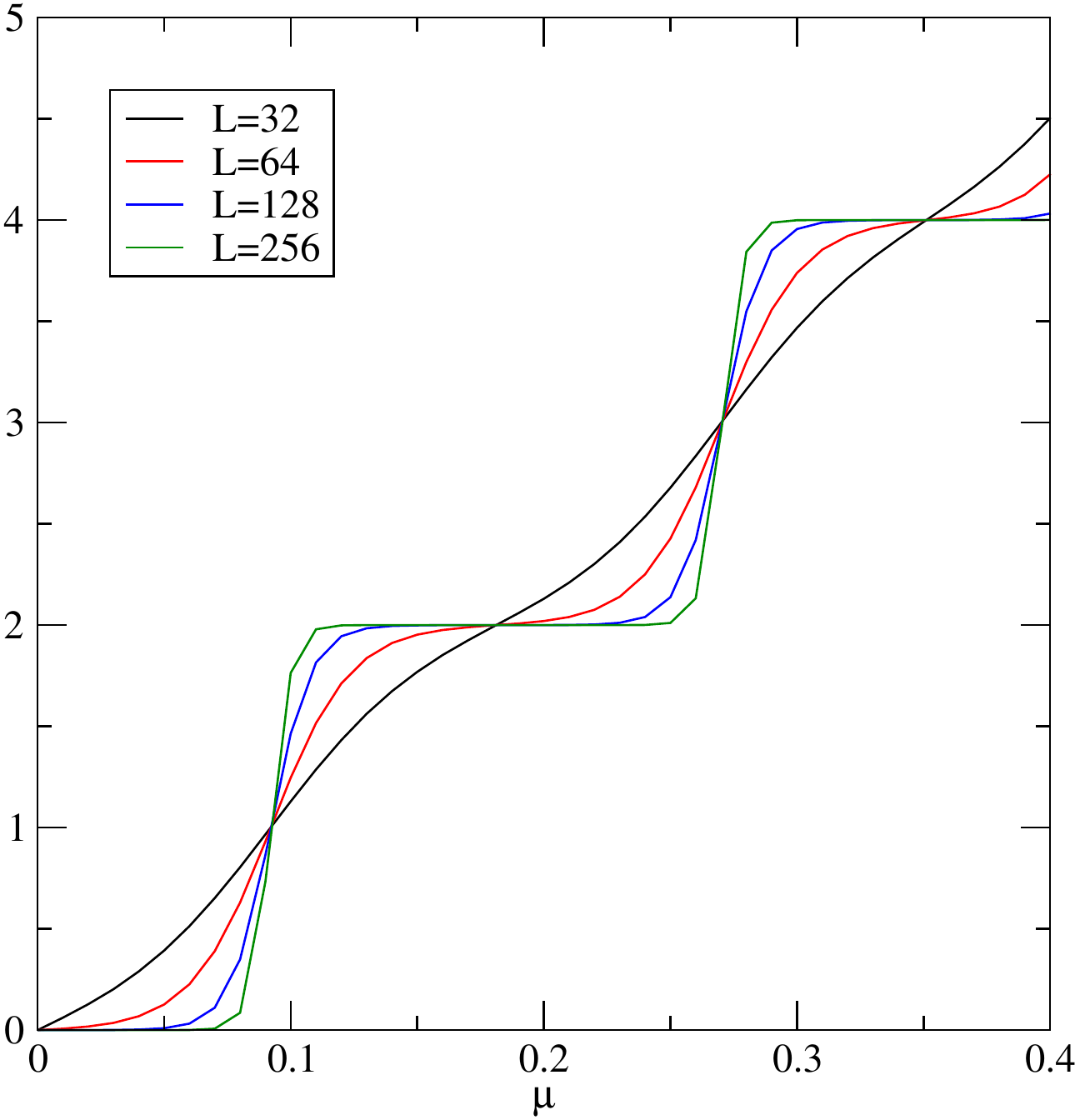}
\includegraphics[height=0.25\linewidth]{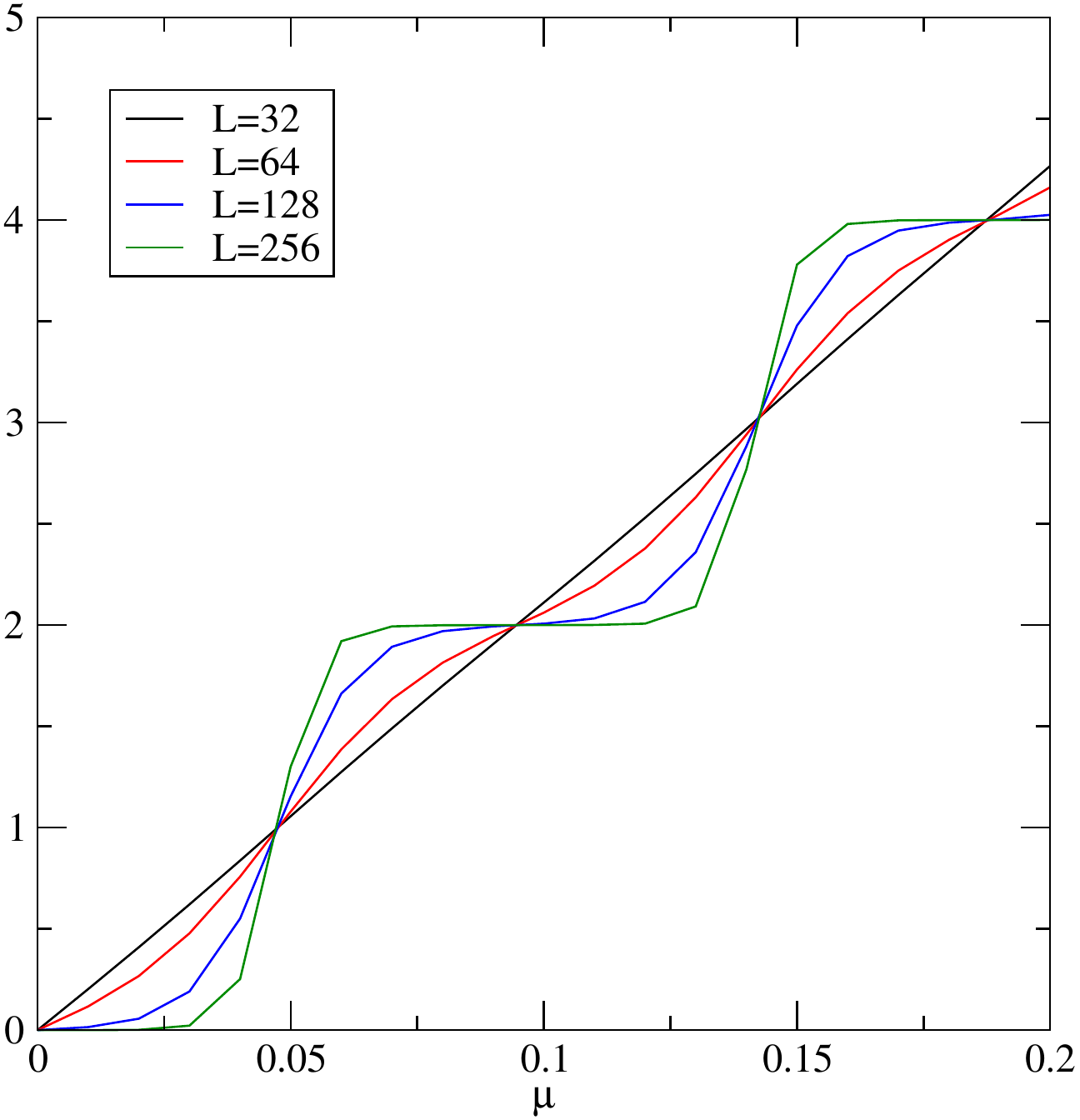}
\caption{ The fermion number $\langle N \rangle$ with open boundary conditions at $U=0$ as a function of the chemical potential. From left to right, $L_X = 12$, $16$ and $32$. }
\label{open_anisotropic_free}
\end{figure*}

\begin{figure*}[t]
\includegraphics[height=0.25\linewidth]{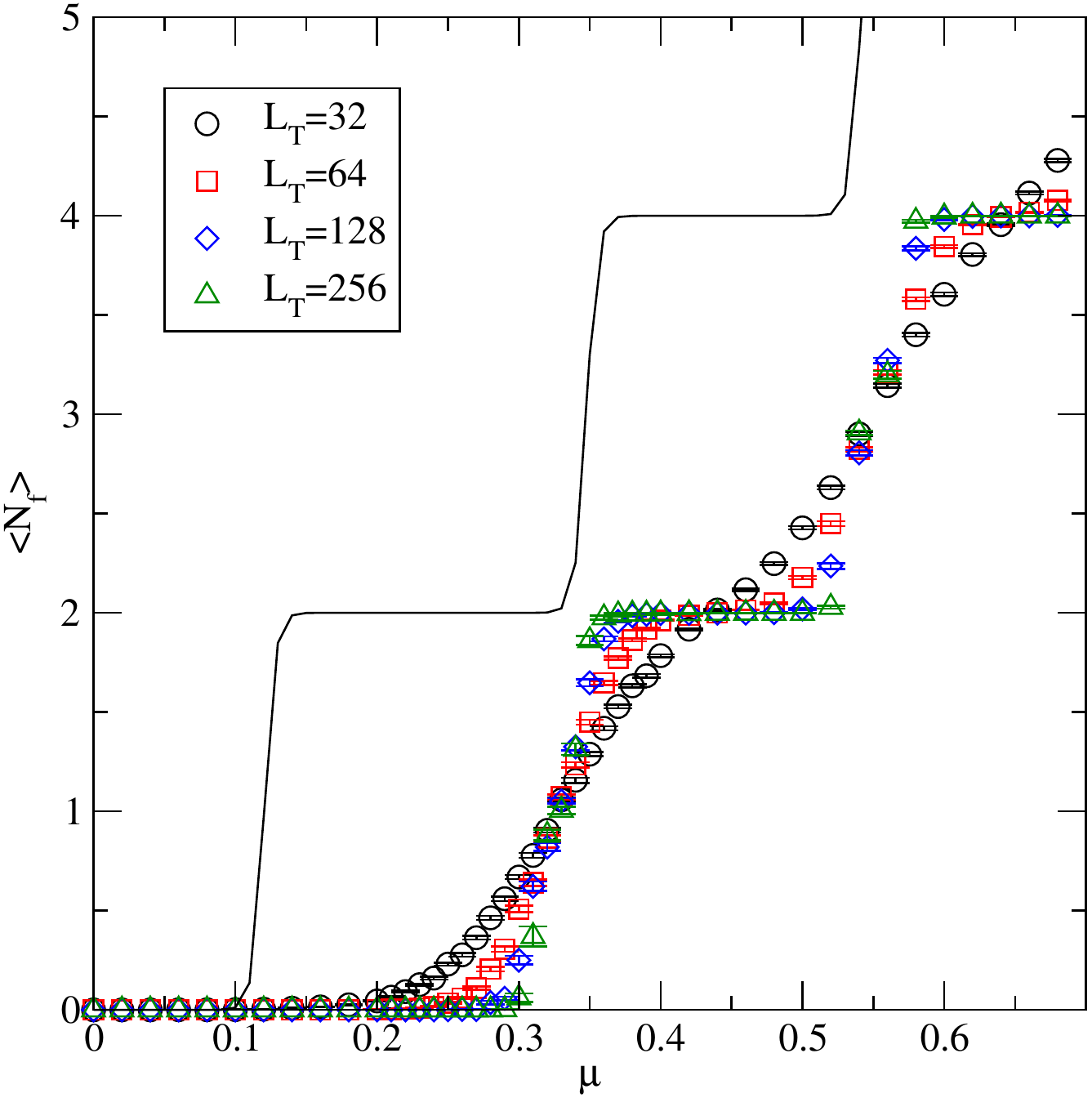}
\includegraphics[height=0.25\linewidth]{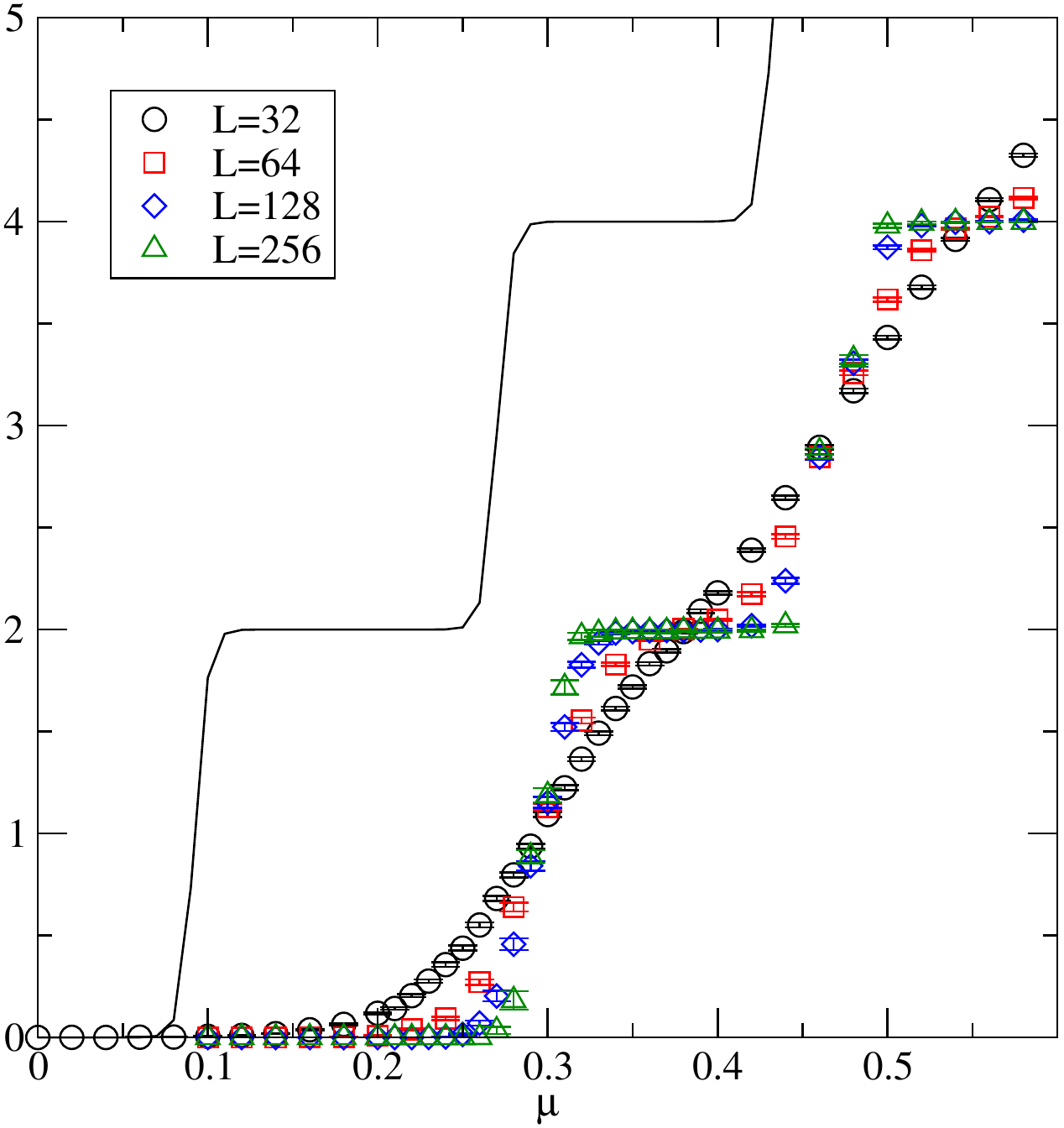}
\includegraphics[height=0.25\linewidth]{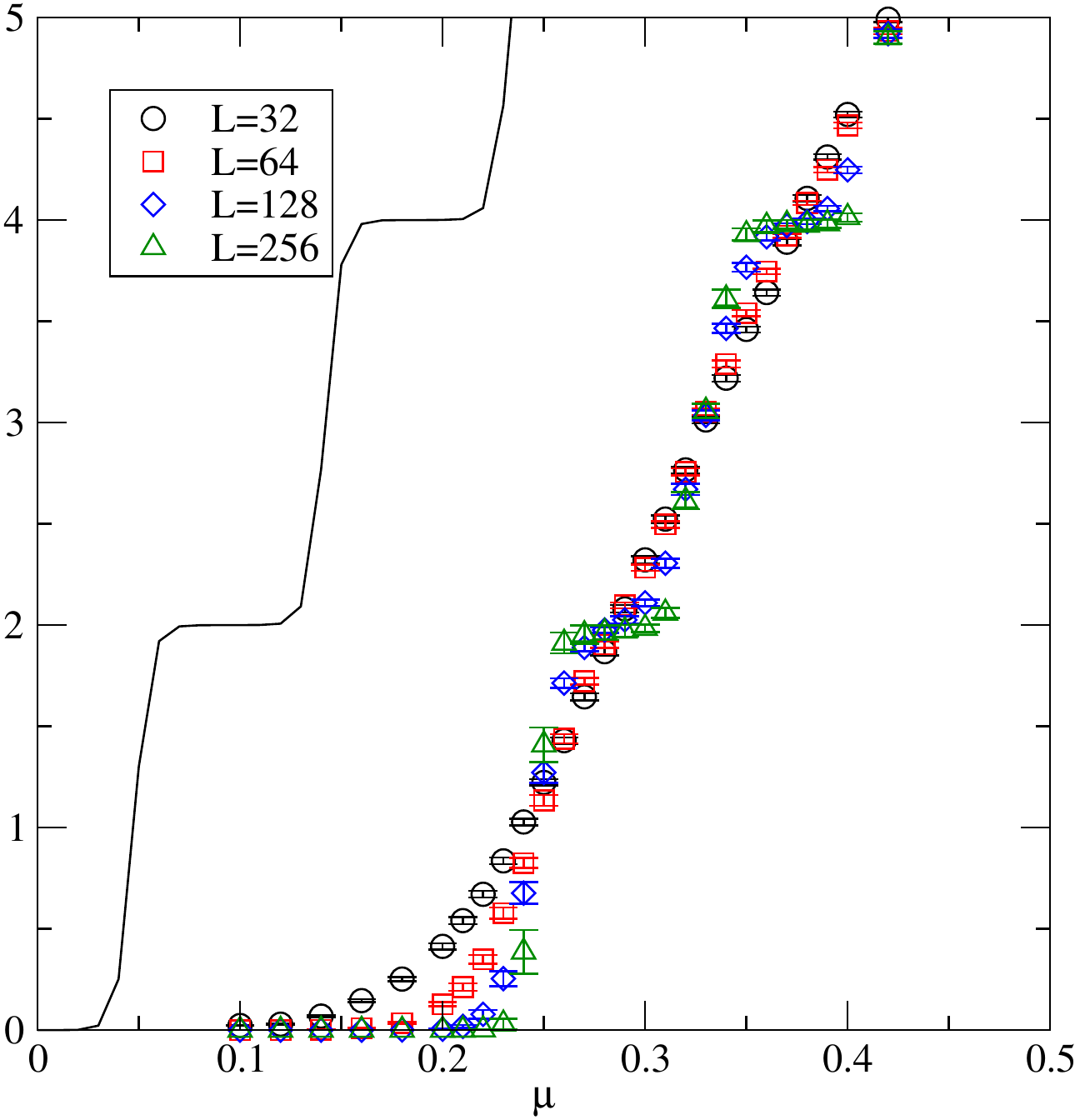}
\caption{ The fermion number $\langle N \rangle$ with open boundary conditions at $U=0.3$ as a function of the chemical potential. The solid line shows the behavior at $U=0$ and $L_T=256$. From left to right, $L_X = 12$, $16$ and $32$. }
\label{open_anisotropic}
\end{figure*}

Next we turn to the physics of finite chemical potential. We first consider a small spatial lattice of $L_X=6$ and $L_T=48$ and study the sign problem in the traditional auxiliary field approach with periodic boundary conditions and compare it with the sign problem in the fermion bag approach with both periodic and anti-periodic boundary conditions. In Fig.~\ref{sign-boundaries} we plot the average sign as a function of the chemical potential in the auxiliary field approach at $U=0.3$ (left) and compare it with that of the fermion bag approach (right). We first wish to learn where the sign problem becomes severe. In the auxiliary field approach the sign becomes severe around $\mu\approx 0.4$, while in the fermion bag approach with anti-periodic boundary conditions it becomes severe around $\mu \approx 0.55$. In the fermion bag approach with periodic boundary conditions the sign problem is never severe, although it is enhanced both at $\mu \approx 0.4$ and then again at $\mu \approx 0.9$. Can we correlate this behavior with some physics?

\begin{table}
\begin{tabular}{| c | c || c | c || c | c || c | c |}
\hline
$\mu$ & $\langle N_f \rangle$  & $\mu$ & $\langle N_f \rangle$ & 
$\mu$ & $\langle N_f \rangle$ & $\mu$ & $\langle N_f \rangle$ \\ 
\hline
\multicolumn{8}{|c|}{Periodic} \\
\hline
0.30 & 0(0) & 0.36 & 0.13(1) &
0.38 & 0.67(2) & 0.40 & 1.52(2) \\
0.42 & 1.91(1) & 0.50 & 2.0(0) &
0.90 & 2.36(3)  & 0.92 & 2.87(2) \\
0.94 & 3.68(3) & 0.95 & 4.03(3) &
0.96 & 4.43(3)  & 0.97 & 4.86(2)  \\
0.98 & 5.21(2) & 0.99 & 5.47(2) &
1.00 & 5.65(2) & 1.10 & 5.98(2) \\
\hline
\multicolumn{8}{|c|}{Anti-periodic} \\
\hline
0.50 & 0.03(3) & 0.54 & 0.19(3) &
0.56 & 0.56(6) & 0.58 & 1.17(5) \\
0.60 & 1.61(9) & 0.61 & 1.71(5) &
0.62 & 1.86(5) & 0.63 & 1.90(4) \\ 
0.64 & 1.86(4) & 0.65 & 1.97(2) & 
0.66 & 1.97(2) & 0.67 & 2.07(2) \\ 
0.68 & 2.04(1) & 0.69 & 2.11(1) & 
0.70 & 2.19(2) & 0.72 & 2.50(3) \\ 
0.80 & 3.95(2) & 0.90 & 4.00(0) & 
1.00 & 4.27(5) & 1.10 & 5.94(2)  \\
\hline
\multicolumn{8}{|c|}{Open} \\
\hline
0.20 & 0.000(0) & 0.34 & 0.044(3) &
0.36 & 0.122(5) & 0.38 & 0.256(7) \\
0.42 & 0.755(8) & 0.44 & 0.974(8) &
0.46 & 1.218(7) & 0.48 & 1.468(8) \\
0.50 & 1.715(8) & 0.52 & 1.869(5) &
0.54 & 1.941(3) & 0.56 & 1.977(2) \\
0.58 & 1.992(1) & 0.60 & 1.996(1) &
0.62 & 2.000(0) & 0.66 & 2.008(1) \\
0.68 & 2.022(2) & 0.70 & 2.056(3) &
0.72 & 2.125(5) & 0.74 & 2.290(8) \\
0.80 & 3.298(6) & 0.82 & 3.600(9) &
0.84 & 3.825(6) & 0.86 & 3.934(4) \\
0.88 & 3.992(2) & 0.90 & 4.041(3) &
0.92 & 4.117(6) & 1.02 & 5.833(7) \\
1.04 & 5.937(4) & 1.06 & 5.975(2) &
1.08 & 5.991(1) & 1.14 & 6.000(0) \\
\hline
\end{tabular}
\caption{ The average fermion number $\langle N_f \rangle$ computed at $U=0.3$ with periodic, anti-periodic and open boundary conditions with $L_X=6$ and $L_T=48$.}
\label{N6-48-table}
\end{table}

\begin{table}[hbt]
\begin{tabular}{|l|c||l|c||l|c||l|c|}
\hline
$\mu$ & $\langle N_f \rangle $  & $\mu$ & $\langle N_f \rangle $  & $\mu$ & $\langle N_f \rangle $  & $\mu$ & $\langle N_f \rangle $  \\
\hline
\multicolumn{8}{|c|}{$L_T=32$}\\
\hline
 0.26 &            $0.28(1)$ &  0.33 &             $1.06(1)$ &  0.42 &            $1.917(6)$ &  0.54 &             $2.90(1)$ \\
 0.28 &           $0.46(1)$ &  0.34 &             $1.16(1)$ &  0.44 &            $2.014(5)$ &  0.56 &             $3.14(1)$ \\
 0.29 &             $0.56(1)$ &  0.36 &           $1.419(10)$ &  0.46 &            $2.116(5)$ &  0.58 &            $3.401(9)$ \\
  0.30 &             $0.67(1)$ &  0.37 &            $1.528(9)$ &  0.48 &            $2.248(7)$ &   0.60 &            $3.604(8)$ \\
 0.31 &             $0.78(1)$ &  0.38 &            $1.632(8)$ &   0.50 &            $2.428(8)$ &  0.62 &            $3.805(7)$ \\
 0.32 &             $0.90(1)$ &  0.39 &            $1.683(9)$ &  0.52 &           $2.631(10)$ &  0.64 &            $3.955(6)$ \\
\hline
\multicolumn{8}{|c|}{$L_T=64$}\\
\hline
 0.26 &            $0.059(5)$ &  0.33 &             $1.07(1)$ &  0.42 &            $1.988(2)$ &  0.54 &             $2.82(1)$ \\
 0.28 &             $0.20(1)$ &  0.34 &             $1.23(1)$ &  0.44 &            $2.000(1)$ &  0.56 &             $3.21(1)$ \\
 0.29 &             $0.31(1)$ &  0.36 &             $1.65(1)$ &  0.46 &            $2.014(2)$ &  0.58 &            $3.580(9)$ \\
  0.30 &             $0.51(2)$ &  0.37 &            $1.772(9)$ &  0.48 &            $2.048(4)$ &   0.60 &            $3.844(7)$ \\
 0.31 &             $0.64(2)$ &  0.38 &            $1.863(7)$ &   0.50 &            $2.178(8)$ &  0.62 &            $3.956(3)$ \\
 0.32 &             $0.86(1)$ &  0.39 &            $1.918(6)$ &  0.52 &             $2.45(1)$ &  0.64 &            $3.994(2)$ \\
\hline
\multicolumn{8}{|c|}{$L_T=128$}\\
\hline
 0.26 &           $0.002(1)$ &  0.33 &            $1.059(5)$ &  0.44 &            $1.999(1)$ &  0.54 &             $2.81(1)$ \\
 0.28 &             $0.04(1)$ &  0.34 &            $1.302(5)$ &  0.46 &            $1.999(1)$ &  0.56 &             $3.27(1)$ \\
 0.29 &            $0.06(1)$ &  0.36 &             $1.87(1)$ &  0.48 &            $2.001(1)$ &  0.58 &             $3.84(1)$ \\
  0.30 &             $0.25(2)$ &  0.37 &            $1.957(6)$ &   0.50 &            $2.019(3)$ &   0.60 &            $3.98(1)$ \\
 0.32 &            $0.84(1)$ &  0.39 &            $1.992(4)$ &  0.52 &             $2.24(1)$ &  0.64 &            $4.0$ \\
\bottomrule
\hline
\multicolumn{8}{|c|}{$L_T=256$}\\
\hline
 0.26 &                 $0.00$ &  0.31 &             $0.42(5)$ &  0.36 &             $2.00$ &  0.54 &             $2.91(1)$ \\
  0.28 &           $0.00$ &  0.32 &             $0.86(2)$ &   0.40 &            $2.00$ &  0.56 &             $3.20(2)$ \\
 0.29 &            $0.00$ &  0.33 &             $1.02(1)$ &   0.50 &            $2.00$ &  0.58 &            $3.97(1)$ \\
  0.30 &             $0.06(2)$ &  0.34 &             $1.31(7)$ &  0.52 &            $2.03(1)$ &  0.64 &           $4.00$ \\
\hline
\end{tabular}
\caption{Monte Carlo results for $\langle N_f\rangle$ at $U=0.3$ with open boundaries at selected values of $\mu$ and $L_T$ for $L_X=12$. This data is plotted in Fig.~\ref{open_anisotropic}.}
\label{LX12}
\end{table}

\begin{table}[htb]
\begin{tabular}{|l|c||l|c||l|c||l|c|}
\hline
$\mu$ & $\langle N_f \rangle $  & $\mu$ & $\langle N_f \rangle $  & $\mu$ & $\langle N_f \rangle $  & $\mu$ & $\langle N_f \rangle $  \\
\hline
\multicolumn{8}{|c|}{$L_T=32$}\\
\hline
 0.21 &            $0.14(1)$ &  0.28 &             $0.80(1)$ &  0.34 &            $1.61(1)$ &  0.44 &           $2.65(1)$ \\
 0.22 &            $0.21(1)$ &  0.29 &             $0.94(1)$ &  0.36 &            $1.83(1)$ &  0.46 &           $2.89(1)$ \\
 0.23 &            $0.28(1)$ &   0.30 &             $1.09(1)$ &  0.37 &            $1.90(1)$ &  0.48 &             $3.17(1)$ \\
 0.24 &             $0.36(1)$ &  0.31 &             $1.22(1)$ &  0.38 &            $1.99(1)$ &   0.50 &            $3.43(1)$ \\
 0.25 &             $0.44(1)$ &  0.32 &             $1.37(1)$ &  0.39 &            $2.09(1)$ &  0.52 &            $3.68(1)$ \\
 0.26 &             $0.55(1)$ &  0.33 &           $1.49(1)$ &  0.42 &            $2.39(1)$ &  0.54 &            $3.91(1)$ \\
\hline
\multicolumn{8}{|c|}{$L_T=64$}\\
\hline
 0.22 &            $0.04(1)$ &   0.30 &             $1.13(2)$ &  0.38 &            $2.002(3)$ &  0.48 &             $3.26(1)$ \\
 0.24 &            $0.09(1)$ &  0.32 &             $1.55(1)$ &  0.42 &            $2.17(1)$ &   0.50 &             $3.62(1)$ \\
 0.26 &             $0.27(1)$ &  0.34 &            $1.83(1)$ &  0.44 &             $2.46(1)$ &  0.52 &            $3.86(1)$ \\
 0.28 &             $0.64(2)$ &  0.36 &            $1.950(5)$ &  0.46 &             $2.85(1)$ &  0.54 &            $3.97(1)$ \\
\hline
\multicolumn{8}{|c|}{$L_T=128$}\\
\hline
 0.23 &           $0.00$ &  0.29 &             $0.84(2)$ &  0.34 &            $1.98(1)$ &  0.46 &             $2.85(1)$ \\
 0.26 &             $0.07(1)$ &   0.30 &             $1.15(3)$ &  0.35 &            $1.99(1)$ &  0.48 &             $3.31(2)$ \\
 0.27 &             $0.20(3)$ &  0.31 &             $1.52(2)$ &   0.40 &            $2.00$ &   0.50 &           $3.88(1)$ \\
 0.28 &             $0.46(3)$ &  0.32 &             $1.83(1)$ &  0.44 &             $2.24(1)$ &  0.54 &            $4.00$ \\
\hline
\multicolumn{8}{|c|}{$L_T=256$}\\
\hline
 0.25 &                 $0.0$ &   0.30 &             $1.19(4)$ &  0.42 &            $2.00(1)$ &  0.48 &             $3.32(2)$ \\
 0.28 &             $0.18(5)$ &  0.31 &             $1.72(4)$ &  0.44 &            $2.02(1)$ &   0.50 &           $3.98(1)$ \\
 0.29 &             $0.89(3)$ &  0.34 &             $1.99(1)$ &  0.46 &             $2.87(2)$ &  0.52 &            $3.99(1)$ \\
\hline
\end{tabular}
\caption{Monte Carlo results for $\langle N_f\rangle$ at $U=0.3$ with open boundaries at selected values of $\mu$ and $L_T$ for $L_X=16$. This data is plotted in Fig.~\ref{open_anisotropic}.}
\label{LX16}
\end{table}

\begin{table}[!t]
\begin{tabular}{|l|c||l|c||l|c||l|c|}
\hline
$\mu$ & $\langle N_f \rangle $  & $\mu$ & $\langle N_f \rangle $  & $\mu$ & $\langle N_f \rangle $  & $\mu$ & $\langle N_f \rangle $  \\
\hline
\multicolumn{8}{|c|}{$L_T=32$}\\
\hline
 0.22 &             $0.67(2)$ &  0.28 &             $1.87(2)$ &  0.36 &             $3.64(2)$ &  0.44 &             $5.40(2)$ \\
 0.24 &             $1.03(2)$ &   0.3 &             $2.32(2)$ &  0.38 &             $4.11(2)$ &  0.46 &             $5.83(1)$ \\
 0.26 &             $1.43(2)$ &  0.34 &             $3.22(2)$ &   0.4 &             $4.52(1)$ &  0.48 &             $6.29(1)$ \\
\hline
\multicolumn{8}{|c|}{$L_T=64$}\\
\hline
 0.21 &             $0.21(2)$ &  0.26 &             $1.44(2)$ &  0.33 &             $3.05(2)$ &   0.4 &             $4.47(1)$ \\
 0.22 &             $0.35(2)$ &  0.28 &             $1.90(2)$ &  0.35 &             $3.54(2)$ &  0.42 &             $4.93(1)$ \\
 0.23 &             $0.58(3)$ &  0.29 &             $2.10(2)$ &  0.36 &             $3.75(1)$ &  0.44 &             $5.43(1)$ \\
 0.24 &             $0.82(2)$ &  0.31 &             $2.50(2)$ &  0.37 &             $3.92(1)$ &  0.46 &             $5.85(1)$ \\
 0.25 &             $1.13(3)$ &  0.32 &             $2.76(2)$ &  0.39 &             $4.25(1)$ &  0.48 &             $6.21(1)$ \\
\hline
\multicolumn{8}{|c|}{$L_T=128$}\\
\hline
 0.21 &            $0.019(6)$ &  0.26 &             $1.71(2)$ &  0.33 &             $3.04(3)$ &   0.4 &             $4.25(2)$ \\
 0.22 &             $0.08(2)$ &  0.28 &             $1.98(1)$ &  0.35 &             $3.77(2)$ &  0.42 &             $4.92(2)$ \\
 0.23 &             $0.25(4)$ &  0.29 &             $2.03(2)$ &  0.36 &             $3.92(2)$ &  0.44 &             $5.64(2)$ \\
 0.24 &             $0.68(5)$ &  0.31 &             $2.31(2)$ &  0.37 &             $3.97(1)$ &  0.46 &             $5.96(1)$ \\
 0.25 &             $1.27(5)$ &  0.32 &             $2.67(3)$ &  0.39 &             $4.06(1)$ &  0.48 &            $6.057(9)$ \\
\hline
\multicolumn{8}{|c|}{$L_T=256$}\\
\hline
 0.25 &             $1.41(9)$ &  0.28 &             $1.96(4)$ &  0.33 &             $3.06(4)$ &  0.42 &             $4.90(3)$ \\
 0.26 &             $1.91(5)$ &  0.31 &             $2.06(2)$ &  0.36 &             $3.97(3)$ &  0.44 &             $5.87(3)$ \\
 0.27 &             $1.95(4)$ &  0.32 &             $2.61(4)$ &   0.4 &             $4.01(2)$ &  0.48 &             $5.99(1)$ \\
\hline
\end{tabular}
\caption{Monte Carlo results for $\langle N_f\rangle$ at $U=0.3$ with open boundaries at selected values of $\mu$ and $L_T$ for $L_X=32$. This data is plotted in Fig.~\ref{open_anisotropic}.
\label{LX32}}
\end{table}

Let us now explore how the fermion chemical potential ``dopes'' the system with fermions. We again focus first on a small lattice, $L_X=6$, $L_T=48$ at $U=0.3$. In table \ref{N6-48-table} we present all of our results for the total fermion number as a function of the chemical potential for open(left), anti-periodic(center) and periodic boundary conditions(right). In Fig.~\ref{nf-boundaries} we plot these results along with the results for free fermions as solid lines. Due to the flavor degeneracy of staggered fermions we expect all states to be at least doubly degenerate. With open periodic boundary conditions this means all jumps must be in steps of two. This is what is observed. With periodic and anti-periodic boundary conditions there is a symmetry between left and right moving particles. With periodic boundary conditions a zero momentum state is allowed which is non-degenerate, hence the first jump in $\langle N \rangle$ near $\mu \approx 0.4$ is only by two. However, the second jump near $\mu \approx 0.9$ is by a factor of four since now non-zero momentum states are excited and each state is doubly degenerate due to the two fermion flavors.  With anti-period boundary conditions the lowest energy state already has momentum and hence again should have four-fold degeneracy. This is clearly seen as a jump of four in the free theory around $\mu \approx 0.5$.  Surprisingly, in the interacting theory this degeneracy of the lowest energy state seems to be broken. We attribute this to the fact that bound state bosons with zero momentum can emerge. The next momentum state is non-degenerate for $L_X=6$ since effectively the lattice size is halved for staggered fermions. This remains unchanged for the interacting theory as well and two additional states are added when $\mu > 1$.

Note that the first step to $\langle N_f\rangle =2$ for both open and periodic boundary conditions occurs around $\mu\approx 0.4$. This coincides with the point where the sign problem becomes severe in the auxiliary field approach, and is somewhat enhanced in the fermion bag approach. The sign problem in the fermion bag approach disappears for large values of $\mu$ until around $\mu \approx 0.9$ where there is the second jump of four in the periodic case. The sign problem in the auxiliary field approach on the other hand never recovers. In the case of anti-periodic boundary conditions the severity of the sign problem coincides with the additional plateau at $\langle N_f\rangle = 2$ which is absent in the free theory as discussed above. While these correlations between sign problems and the underlying physics are not surprising, the fact that energies and degeneracies of the lowest lying states can be influenced by boundary conditions and interactions on small lattices offers an excellent opportunity for methods that claim to solve the sign problems to reproduce them. 

Since the sign problem is absent with open boundary conditions we can use it to study the behavior of $\langle N_f\rangle$ on large asymmetric lattices ($L_X \neq L_T$) so as to understand the physics of fermion doping at a fixed $L_X$. One of the main results that our model shares with QCD is that fermions become massive entirely due to interaction effects and the value of the chemical potential where the first jump in $\langle N_f\rangle$ occurs will be this finite size fermion mass $m_f^{L_X}$. In order to see the effects of interactions we plot $\langle N_f\rangle$ as a function of $\mu$ in the free theory (Fig.~\ref{open_anisotropic_free}) and in the interacting theory with $U=0.3$ (Fig.~\ref{open_anisotropic}) both with open boundary conditions. Selected data points have also been tabulated in Tables.~\ref{LX12},\ref{LX16} and \ref{LX32} for benchmark purposes.

We study three different lattice sizes $L_X=12$ (left) $L_X=16$ (center) and $L_X=32$ (right). For each of these lattices we study the effects of increasing $L_T$. Note that the critical value of $\mu$ where the first jump to $\langle N_f\rangle =2$ occurs, shifts to lower values as $L_X$ increases in the free theory. We expect this value to vanish in the large $L_X$ limit since fermions are massless. However, in the interacting theory we note the jump change in the critical value is smaller and should approach $0.183(1)$ (see table \ref{table_large_volume}) as $L_x$ becomes large. Also the jump becomes sharper as the anisotropy (value of $L_T$) is increased and approaches a step function as expected. To quantify the value of $m_f^{L_X}$ we measure $\langle N_f\rangle$ for several values $\mu$ near the transition at two different values of $L_T$.  In particular with $L_X=12$ we use $L_T=64,128$ and with $L_X=32$ we use $L_T=128,256$. We find the value of $\mu$ where $\langle N_f\rangle$  measured with different $L_T$'s cross using a linear fit near the crossing. These values of $\mu$ are taken to be estimates of $m_f^{12}$ and $m_f^{32}$. These numbers for different values of $U$ are tabulated in table \ref{table_mcrit_fit}. Similarly by fitting the chiral condensate susceptibility to the form
\begin{equation}
\chi = \chi_0 + B \mathrm{e}^{-m_b^{L_X} L_T}
\end{equation}
we can also extract the finite size boson mass $m_b^{L_X}$. These values are also given in table \ref{table_mcrit_fit} for $L_X=12$ and $32$. We find that while $m_f^{L_X}$ increases sharply with $U$, $m_b^{L_X}$ decreases mildly.

\begin{table}[tbh]
\center
\begin{tabular}{|c|c|c|c|c|}
\hline
$U$ & $m_b^{12}$ & $m_f^{12}$ & $m_b^{32}$ & $m_f^{32}$  \\ 
\hline
0    & 0.17207  & 0.120(1) & 0.067393 & 0.045(5)  \\
0.1  & 0.158(6)  & 0.163(2) & 0.061(4) & 0.0705(6) \\
0.2  & 0.184(6) & 0.235(10) & 0.06(1)  & 0.1397(3) \\
0.3  & 0.156(4) & 0.328(2) & 0.066(7) & 0.247(2)  \\
0.4  & 0.143(4) & 0.425(2) & 0.060(3) & 0.356(1)  \\
0.5  & 0.143(7) & 0.519(2) & 0.057(2) & 0.465(1) \\
0.6  & 0.137(1) & 0.601(2) & 0.049(5) & 0.556(1) \\
1.0  & 0.121(4) & 0.871(1) & 0.050(4) & 0.842(2)  \\
$\infty$ & 0.114(4) & $\infty$ & 0.0476(9)& $\infty$ \\
\hline
\end{tabular}
\caption{The fermion and boson masses measured using the fermion number with open boundary conditions. The boson mass at $U=0$ is calculated directly from the free correlator on a finite lattice.  }
\label{table_mcrit_fit}
\end{table}

\section{Conclusions} \label{conclusions}

In this work we have studied the $1+1$ dimensional lattice Thirring model with staggered fermions at both zero and finite densities. We showed that the model is free of sign problems in the massless limit when open boundary conditions are used. In this case we used the worldline formulation to study the model. In the case of periodic and anti-periodic spatial boundary conditions the sign problem is mild on square lattices but becomes severe when on asymmetric lattices. However, the fermion bag formulation seems to alleviate the problem except at critical values of the chemical potential where fermion number jumps. We provide accurate estimates for the total particle number as a function of the chemical potential for a few lattice sizes. Our results could be used as a benchmark for future studies by other methods that attempt to solve the sign problem.

\section*{Acknowledgments}

We thank A.~Alexandru and P.~Bedaque for extensive discussions about their work and for providing their results so we can compare against our results on small lattices where such a comparison was possible. SC and JR's work was supported by the U.S. Department of Energy, Office of Science, Nuclear Physics program under Award Number DE-FG02-05ER41368. VA's work was supported by the U.S. Department of Energy under grant number DE-SC0010005 .

\bibliography{ref}

\end{document}